\documentclass[11pt,a4paper]{article}
\usepackage{amssymb,amsfonts,amsmath,amsthm}
\usepackage[english]{babel}
\usepackage[left=2.7cm,right=2.7cm,bottom=2.67cm,top=2.67cm]{geometry}
\usepackage{dsfont}
\usepackage{mathrsfs}
\usepackage{natbib}
\usepackage{enumerate}
\usepackage{color}
\usepackage[ruled]{algorithm2e}
\usepackage[shortlabels]{enumitem}
\usepackage{graphicx}
\usepackage{multirow}
\usepackage{multicol}
\usepackage{subfigure}
\usepackage{placeins}
\usepackage{diagbox}
\usepackage{setspace}
\usepackage{url}
\usepackage{arydshln} 
\graphicspath{{figures/}}
\usepackage[final]{pdfpages}
\theoremstyle{plain}
\newtheorem{thm}{Theorem}[section]

\theoremstyle{definition}
\newtheorem{rmk}{Remark}[section]

\newcommand{\E}{\mathds{E}}
\newcommand{\F}{\mathscr{F}}
\newcommand{\R}{\mathds{R}}

\newcommand{\Z}{\mathds{Z}}

\newcommand{\bs}[1]{\boldsymbol{#1}}
\newcommand{\hbs}[1]{\hat{\boldsymbol{#1}}}

\newcommand{\var}{\mathrm{Var}}

\long\def\sfootnote[#1]#2{\begingroup%
\def\thefootnote{\fnsymbol{footnote}}\footnote[#1]{#2}\endgroup}

\def\bfootnote{\xdef\@thefnmark{}\@footnotetext}

\graphicspath{{figures/}}

\begin{document}
\pagestyle{myheadings} 
\markboth{Multiple imputation for observation-driven models}{G. Pumi, T.S. Prass and D.K. Verdum}

\thispagestyle{empty}
{\centering
\Large{\bf A Novel Multiple Imputation Approach For Parameter Estimation in Observation-Driven Time Series Models With Missing Data }\vspace{.5cm}\\
\normalsize{ {\bf Guilherme Pumi${}^{\mathrm{a,}}$\sfootnote[1]{Corresponding author. This Version: \today},\let\thefootnote\relax\footnote{\hskip-.3cm$\phantom{s}^\mathrm{a}$Mathematics and Statistics Institute and Graduate Program in Statistics - Universidade Federal do Rio Grande do Sul.
}  Taiane Schaedler Prass${}^\mathrm{a}$ and Douglas Krauthein Verdum${}^\mathrm{a}$
 \\
\let\thefootnote\relax\footnote{E-mails: guilherme.pumi@ufrgs.br (G. Pumi), taiane.prass@ufrgs.br (T.S. Prass), dkverdum@gmail.com (D.K. Verdum) }
\let\thefootnote\relax\footnote{ORCIDs: 0000-0002-6256-3170 (Pumi); 0000-0003-3136-909X (Prass); 0009-0009-4136-6731 (Verdum).}}\\
\vskip.3cm
}}

\begin{abstract}
In this paper we propose a novel multiple imputation method specifically designed for parameter estimation in observation-driven models (ODM). The approach takes advantage of the iterative nature of the systematic component in ODM to propagate the dependence structure through missing data, thereby minimizing its impact on estimation. Unlike traditional imputation techniques, the proposed method accommodates continuous, discrete, and mixed-type data while preserving key distributional and dependence properties. Under general conditions, we prove that the proposed method converges almost surely to the true parameter, regardless of the missing mechanism or the proportion of missing data. We evaluate its performance through Monte Carlo simulations in the context of GARMA models, considering time series with up to 70\% missing data. Two practical applications to environmelntal time series further demonstrates its practical utility.
\vspace{.2cm}

\vspace{.2cm}

\end{abstract}
\section{Introduction}

The literature on missing data is extensive, with numerous books covering a wide range of methods, primarily in the context of independent data \citep[e.g.,][]{littleandrubin, buuren, carp}. As noted in \cite{buuren}, relatively little is known about imputation for time series. Unlike independent data, handling missing values in time series is significantly more challenging due to temporal dependence. There are two primary approaches to dealing with missing data in time series when inference and forecasting are of interest. The first, imputation, is the simplest and most widely used. It involves replacing missing values with plausible estimates and proceeding with analysis as if no data were missing. The second approach relies on estimators specifically designed to handle missing data without requiring a fully completed time series. When available, these estimators typically yield better results, but they are difficult to derive and often highly model-dependent.

This work focuses on imputation in the context where missing data mechanism is completely random. Various methods exist, with many implemented in software packages. For instance, the R package \texttt{imputeTS} \citep{imputets} provides 18 general-purpose imputation methods, including mean, median, and mode imputation, last observation carried forward, next observation carried backward, spline interpolation, and random imputation. These methods are broadly applicable, but they may fail in specific contexts. For example, in time series constrained to the unit interval $(0,1)$, spline interpolation often generate imputed values outside the boundaries, making it unsuitable for GARMA models for double-bounded data.

General-purpose imputation methods have the advantage of being quick and simple, which adds to its appeal, but they are not without drawbacks. First, they tend to underestimate variance, introducing biases in parameters that depend on it, such as correlations. Mean or median substitution, for instance, replaces distinct missing values with a single fixed number, artificially reducing variance and introducing a point mass in an otherwise continuous distribution. Second, imputed values alter the dependence structure of the time series in complex and often unpredictable ways, which is problematic since many statistical models rely on accurately capturing this structure. Third, standard errors and $p$-values are distorted, as standard inferential procedures assume that all observed values are accurate, failing to account for the additional uncertainty introduced by imputation. As the proportion of missing values increases, these issues worsen, requiring more specialized techniques.

A promising strategy to mitigate these issues is multiple imputation, in which missing values are replaced by plausible random draws from a carefully chosen distribution. This process is repeated $m>1$ times, generating $m$ complete time series. Inference is performed on each time series separately, and results are combined using simple pooling rules to obtain a final estimate with an associated standard error. The advantage of multiple imputation is that it accounts for the additional uncertainty introduced by the imputation process, but it presents two major problems: how to choose the distribution from which the imputed values are to be draw from, and how to do it smartly, aiming at minimizing the aforementioned shortcomings of imputation methods?

A simple but naive multiple imputation approach is to draw missing values uniformly from the observed range of the time series. However, this fails to address the problems outlined earlier. A more refined approach was proposed by \cite{gladys}, in which missing values are drawn from a truncated normal distribution centered at the previous observed value, with variance calibrated to match the sample variance of the observed data. This method aims to preserve the local dependence structure while maintaining variance consistency. However, the results in \cite{gladys} showed only marginal improvements over standard imputation methods, which may be attributed to the specific application context - namely, long-range dependent processes, where the method's local nature may not be sufficient to enhance estimation.

In this work, we introduce a novel multiple imputation method designed specifically for parameter estimation in observation-driven models. Our approach takes advantage of the systematic component’s iterative nature in these models to propagate the conditional dependence structure through missing values, reducing imputation-induced distortions in dependence. The method is applicable to continuous, discrete, or mixed random components and can be used with any estimator in observation-driven models.

Informally, given an initial parameter estimate, missing values are imputed by reconstructing the systematic component of the model and sampling from the assumed conditional distribution. This ensures that the dependence structure of the imputed series closely matches that of the observed data. After each imputation step, model parameters are reestimated using the newly completed dataset. This process is iterated several times, with estimates from multiple imputed datasets combined to produce a final parameter estimate. The method is compatible with any estimator of any quantity of interest in observation-driven models, making it widely applicable. Under general conditions, we show that the proposed method converges almost surely to the true parameter. An interesting feature of our method is that strong consistency is attained regardless of the missing mechanism or the proportion of missing data.

We evaluate the proposed method through a series of Monte Carlo simulation studies in the context of GARMA models, considering time series with up to 70\% missing data. Finally, we illustrate its practical application by analyzing the proportion of stocked energy stored in South Brazil and the PM${}_{2.5}$ pollution data in Brandon, Canada.
\section{Preliminaries}
In this work we consider observation-driven models as discussed in \cite{Cox1981}. More specifically, let $\{Y_t\}_{t\in\Z}$ be a time series of interest and let $\{\bs X_t\}_{t\in\Z}$ denote a set of $r$-dimensional exogenous time dependent (possibly random) covariates. Let $\F_{t}$ denote the information ($\sigma$-field) available to the observer at time $t$, that is,   $\F_{t}:=\sigma\{\bs X_{t+1}, Y_{t},\bs X_{t},Y_{t-1},\cdots\}$, where, by convention, $\bs X_t$ denotes the observed values at time $t$ for deterministic covariates, and at time $t-1$ for stochastic ones. The model's random component is specified assuming that $Y_t$ given $\F_{t-1}$ is distributed according to a distribution depending on an identifiable parameter vector $\bs\nu\in\R^v$ and a quantity of interest $\mu_t$ satisfying
\begin{equation}\label{eod}
g(\mu_t) = A(\bs X_{t}^\prime,Y_{t-1},\bs X_{t-1}^\prime,Y_{t-2},\cdots; \bs\lambda),
\end{equation}
for some twice continuously differentiable with respect to $\bs\lambda$, and integrable real function $A$, where $\bs\lambda$ is an identifiable finite dimensional parameter vector and $g$ is a twice continuously differentiable link function. Observe that, specification \eqref{eod} implies that $g(\mu_t)$ is $\F_{t-1}$-measurable. Let $\bs\gamma=(\bs\nu',\bs\lambda')'$ be the parameter of interest and denote by $f(\cdot; \bs\gamma| \F_{t})$ be the conditional density, mass or density-mass hybrid (mixed type) related to $Y_t$ given $\F_t$. Equation \eqref{eod} determines the model's systematic component. The observation-driven model we are interested in this paper is given by the specification of the random component and the systematic component \eqref{eod}.

One of the main advantages of observation-driven models is that its dynamic is sequentially determined given the model's past knowledge. This allows for sequential conditional inference regarding model parameters without requiring full knowledge of the joint distribution, leading to the partial likelihood approach as discussed in \cite{Cox1975} and \cite{wong1986}  \citep[see also][]{Fokianos2004}.

The full generality presented in \eqref{eod} is sufficient for developing the method proposed and studied in this paper, albeit of limited practical use without a complete specification of the function $A$ and its interaction with parameter $\bs\lambda$. Also, the distributional parameter $\bs\nu$ in most cases is uni-dimensional.  Model \eqref{eod} encompasses many interesting particular cases of interest:
\begin{itemize}
\item By taking
\[A(\bs X_{t}^\prime,Y_{t-1},\bs X_{t-1}^\prime,Y_{t-2},\cdots; \bs\lambda)=\bs X_t'\bs\beta+\sum_{i=1}^p\phi_j\mathcal A(Y_{t-j}, \bs X_{t-j}, \bs \beta)+\sum_{i=1}^q\theta_j\mathcal{M}(Y_{t-j}, \mu_{t-j}),\]
for $\mathcal{A}$ and $\mathcal{M}$ representing the autoregressive and moving average terms with associated parameters $\bs\phi=(\phi_1, \cdots, \phi_p)'$ and $\bs\theta=(\theta_1, \cdots, \theta_q)'$, respectively, and $\bs\beta=(\beta_1, \cdots,\beta_r)'$ is the parameter associated to the covariates. This specification for $g(\mu_t)$ resembles a set of ARMA equations with exogenous covariates, often referred to as ARMAX, but it will only satisfy a set of ARMA recurrence equations if $\mathcal{M}$ is carefully chosen. This general form, briefly mentioned in \cite{Benjamin2003}, is still too general for practical purposes.
\item  A specialization of the previous specification is the base of all the models know in the literature as the Generalized ARMA models \citep{Benjamin2003}, abbreviated GARMA when the random component is a distribution member of the exponential family in canonical form, and called GARMA-like models otherwise, has the form
\begin{equation}\label{arma}
g(\mu_t)=\bs X_t'\bs\beta+\sum_{j=1}^p\phi_j\big(g(Y_{t-j})-\bs X_{t-j}' \bs\beta\big) + \sum_{i=1}^q\theta_ir_{t-i},
\end{equation}
with $r_{t-j}$ taking one of two commonly used forms. In the predictive scale, $r_{t}=g(Y_t)-g(\mu_t)$ \citep[][to cite just a few]{Rocha2009, Bayers, PUMI2019, uw}. With this choice \eqref{eod} can be shown to satisfy an ARMA set of difference equations and the AR and MA parameters have their usual interpretation. However, $\E(r_t)\neq0$ and nothing is known about the distribution of $r_t$; Considering the response scale instead, namely $r_t=Y_t-\mu_t$, the pair $\big\{(r_t,\F_{t})\big\}_{t\in\Z}$ is a martingale difference, provided $\E(Y_t)<\infty$. This fact is often used for goodness-of-fit purposes \citep[see][]{ibarma,helen,ptsrarxiv}.
\item PARX models introduced by \cite{PARX} consider a term depending of exogenous covariates (hence the X in the name) with specification
\[A(\bs X_{t-1}^\prime,Y_{t-1},\bs X_{t-2}^\prime,Y_{t-2},\cdots; \bs\alpha, \bs \beta,\bs\gamma) = \omega+\sum_{i=1}^p\alpha_iY_{t-i}+\sum_{j=1}^q\beta_j\mu_{t-j}+f(\bs X_{t-1},\bs\gamma),\]
where $f(\cdot;\bs\gamma):\R^r\rightarrow[0,\infty)$ is a user chosen function depending on a parameter $\bs\gamma\in\R^g$. The additive specification $f(\bs X_{t-1},\bs\gamma)=\sum_{i=1}^g\gamma_if_i(\bs X_{t-i})$ for real functions $f_i$ is often used in practice.
\item The Beta Autoregressive Chaotic ($\beta$ARC) models, introduced by \cite{BARC}, considers a systematic component that resembles \eqref{arma} but substituting the MA term  by a chaotic process, yielding the specification
\begin{equation*}
g_1(\mu_t)=\bs X_t'\bs\beta+\sum_{j=1}^p\phi_j\big(g_1(Y_{t-j})-\bs X_{t-j}' \bs\beta\big) + g_2\big(\mathrm{T_\theta}^{t-1}(U_0)\big),
\end{equation*}
where $g_1$ and $g_2$ are twice differentiable link functions, $\mathrm T_\theta:[0,1]\rightarrow[0,1]$ is a dynamical system, depending on a parameter $\theta$ and $\mathrm T^t=\mathrm T\circ \mathrm T^{t-1}$ denotes the $t$-fold composition. The dynamical system $\mathrm T_\theta$ is assumed to posses an absolutely continuous invariant measure from which $U_0$ (a random variable) is distributed accordingly. $\beta$ARC models present several attractive and interesting properties, allowing the modeling of many dependence structures that are not possible to achieve with other GARMA structures. See \cite{BARC}, for more details.
\item The  Rayleigh Generalized Autoregressive Score (Ray-GAS) model of \cite{gasarma} considers a random component following the Rayleigh distribution parameterized in terms of its mean $\mu_t$ and systematic component following a GAS structure of the form
\begin{equation*}
\log(\mu_t) = \bs X_t'\bs\beta + \frac18\sum_{i=1}^p \phi_i\bigg(\frac{\pi y_{t-i}}{\mu_{t-i}}-4\mu_{t-i}\bigg) + \sum_{j=1}^q \theta_j\log(\mu_{t-j}).
\end{equation*}
In this case, parameters $\phi_i$ and $\theta_j$ do not have an ARMA-like interpretation.
\end{itemize}

\section{Multiple Imputation Approach for Parameter Estimation.}

In this section we present the proposed multiple imputation approach for parameter estimation in the context of ODMs, considering that missing data occurs  completely at random. Let $Y_1, \cdots, Y_n$ be a sample drawn from a ODM for which the random component is distributed according to $f(\cdot|\bs \gamma_0;\F_{t-1})$, whereas the systematic component follows \eqref{eod}, for  some quantity of interest $\mu_t$. The identifiability of $\bs\lambda$ implies that the dependence of $f(\cdot|\bs \gamma_0;\F_{t-1})$ on $\bs\lambda_0$ is uniquely tied to the specification of $\mu_t$ and conversely. Hence, writing $f(\cdot|\bs \gamma_0;\F_{t-1})$ is equivalent to $f(\cdot|\mu_t, \bs\nu_0;\F_{t-1})$. This notation will prove useful in what follows. Suppose that, given a complete sample from $\{Y_t\}_{t\in\Z}$, a method for estimating  $\bs\gamma$ is available. At this point, no further assumption is required.

Let $\mathcal{I}$ denote the subset of indexes for which $Y_t$ is observed (not-missing). Naturally, its complement, $\mathcal{I}^c$, is the set of indexes for which $Y_t$ is missing. The particular missing data mechanism is irrelevant for our purposes.  The idea of the proposed multiple imputation method for parameter estimation in ODMs is as follows.

Given an initial estimate $\hbs\gamma_0=(\hat{\bs\nu}_0',\hat{\bs\lambda}_0')'$ and $\{Y_t\}_{t\in \mathcal I}$, missing values are imputed sequentially using a random draw from $f(\cdot | \hat\mu_t, \hat{\bs\nu}_0;\F_{t-1})$, where $\hat\mu_t$ is reconstructed using the previous information through
\begin{equation}\label{recon}
    \hat{\mu}_t = g^{-1} \big( A(\bs X_t, Y_{t-1}^\ast, \bs X_{t-1},Y_{t-2}^\ast, \cdots;\hat{\bs\lambda}_0) \big).
\end{equation}
with $Y_t^\ast$ denoting $Y_t$, if $t\in\mathcal I$, or its imputed value, otherwise. This imputation process generates a sample $y_1^1, \cdots,y_n^1$, referred to as the imputed sample, from which we estimate $\hbs\gamma_0^1$.

This procedure is repeated $K-1$ more times, resulting in a set of independently (given $\mathcal{I}^c$) estimated parameters $\hbs\gamma_0^1,\cdots,\hbs\gamma_0^K$. We define $\hbs\gamma_1:=K^{-1}\sum_{k=1}^K\hbs\gamma_0^k$, as the new reference value for $\bs\gamma$. Then, the procedure discussed so far is iterated using $\hbs\gamma_1$ as reference, yielding a new estimate $\hat{\bs\gamma_2}$ in the end, and so on. These steps are repeated until convergence is achieved according to a predetermined criteria, or until a predetermined maximum number of iterations, say $H$, is reached. In any cases, the last estimate obtained, say $\hbs\gamma$, is taken as the final estimated value for $\bs\gamma$. This process is outlined below in the form of an algorithm, which systematically illustrates these steps.

\begin{algorithm}[H]
\vspace{0.3cm}
\begin{enumerate}[leftmargin = 0.5em]
    \item Given the observed time series $\{y_t\}_{t\in \mathcal{I}}$, compute the initial estimate $\hat{\bs{\gamma}}_0$.
    \item\label{it:2} Set $k=1$. Starting from $\hbs\gamma_0$, sequentially impute each missing value $y_t$ for $t\in\mathcal{I}^c$ by sampling from $f(\cdot | \hat{\mu}_t, \hat{\bs\nu}_0)$, where $\hat\mu_t$ is recursively computed using \eqref{recon}. Denote the imputed series by $y_1^1,\dots, y_n^1$.
    \item\label{it:3} Using the imputed series $y_1^1,\cdots, y_n^1$, compute the updated estimate $\bs{\hat\gamma}_0^1$.
    \item\label{it:4} Repeat steps \ref{it:2} and \ref{it:3} for $k\in\{2,\cdots,K\}$, obtaining a set of estimates $\hbs\gamma_0^2,\cdots,\hbs\gamma_0^K$.
    \item\label{it:5} Define $\hat{\bs{\gamma}}_1$ as the component-wise sample mean of the estimates $\hbs\gamma_0^1,\cdots,\hbs\gamma_0^K$.
    \item  Iterate steps \ref{it:2} to \ref{it:5}, updating the estimate at each iteration, until convergence is achieved according to some criteria or until a predefined maximum number of iterations $H$ is reached. In any case, the last estimate, say $\hbs\gamma$ is set as the final estimate.
\end{enumerate}
\caption{Proposed multiple imputation approach for parameter estimation.}
\end{algorithm}

If we let $\hbs\gamma_h^1,\cdots,\hbs\gamma_h^K$ be the estimates from which $\hbs\gamma$ was obtained, we can estimate the uncertainty in the estimation of $\bs\gamma_0$ associated with the data imputation by calculating its standard deviation, for instance.
\section{Theoretical Discussion}\label{TD}
In this section we present a discussion related to the theoretical foundation guaranteeing the almost sure convergence of Algorithm 1 to the true parameter vector $\bs\gamma_0$. To do that, we shall rely on techniques from dynamical system and ergodic theory. The material discussed here can be found in the books by \cite{HK}, \cite{Devaney} and \cite{Walters}. Our goal is to show that, under general conditions, if the algorithm is started using a point chosen from a set of ``good'' initial points (to be make precise later), then Algorithm 1 will almost surely converge to the true parameter value.

Let $\{Y_t\}_{t\in\Z}$ be a stochastic process following an ODM of interest, with finite second moment and true parameter value $\bs\gamma_0\in\Omega\subseteq\R^d$, where $\Omega$ denotes the parameter space. Let $\F_{0}$ represent the set of initial conditions for the model, that is, $\F_0:=\sigma\{Y_0,Y_{-1},\cdots\}$. Let $Y_1,\cdots,Y_n$ be a sample from $\{Y_t\}_{t\in\Z}$. Assume that $m<n$ values are missing and, without loss of generality, that $Y_1$ and $Y_n$ are not missing. We make no assumption on the missing data mechanism. Let $\hat{\bs \gamma}_0\in\Omega$ be a starting point for Algorithm 1, and suppose we are using a strongly consistent estimator which we denote abstractly by $\hbs\gamma$, or by $\hbs\gamma_{s,n}:\Omega\rightarrow\Omega$ when the dependence in $s$ and $n$ is paramount. We assume that the conditions for the consistency of $\hbs\gamma$ are fulfilled.

We will need to make small changes to the notation in Algorithm 1 to make some dependencies more explicit. Recall that, starting with a point $\hbs\gamma_0\in\Omega$ and from a sample of size $n$, after $s$ iterations of Algorithm 1, we arrive at an estimate $\hbs \gamma_{s,n}$ of $\bs\gamma_0$, from which we impute the missing data using Algorithm 1, obtain estimates $\hbs\gamma_{s,n}^{1}, \cdots,\hbs\gamma_{s,n}^{K}$ using $\hbs\gamma$, and set $\hbs\gamma_{s+1,n}=\frac1K\sum_{i=1}^K \hbs\gamma_{s,n}^{i}$. Observe that $\hbs \gamma_{s,n}$ depends on $K$ and $n$, for all $s>0$ whereas $\hbs\gamma_0$ does not. We shall omit the dependence of $\hbs \gamma_{s,n}$ in $K$, since we shall work with an equivalent but more precise notation, as follows. Let $T_{K,n}:\Omega\rightarrow\Omega$ be the measurable transformation that takes $\hbs \gamma_{s,n}$ and return $T_{K,n}(\hbs \gamma_{s,n})=\hbs \gamma_{s+1,n}$. The iterative nature of transformation $T_{K,n}$ allows us to write
\begin{equation}\label{dtsd}
\hbs\gamma_{s+1,n} =T_{K,n}(\hbs\gamma_{s,n})= T_{K,n}\big(T_{K,n}(\hbs\gamma_{s-1,n})\big) = \cdots  = (\underbrace{T_{K,n}\circ\cdots\circ T_{K,n}}_{s+1 \text{ times}})(\hbs\gamma_0) = T_{K,n}^{s+1}(\hbs\gamma_0),
\end{equation}
where, for a transformation $T$, $T^p$ denotes the $p$-fold composition of $T$. Relationship \eqref{dtsd} along with $\Omega$ determines a (discrete-time) dynamical system.   Convergence properties of the proposed algorithm will depend on the ergodic properties of the transformation $T_{K,n}$ and its iterations as well as the choice of $\hbs \gamma_0$.

The relationship $T_{K,n}(\hbs\gamma_{s,n})=T^s_{K,n}(\hbs\gamma_0)$ allows the study of its limiting behavior in $s$ under the point of view of dynamical systems. For a transformation $T:\Omega\rightarrow\Omega$ and $\bs x\in\Omega$, the sequence $\bs x,T(\bs x),T^2(\bs x), \cdots$ is called the orbit of $\bs x$ under $T$. A point $\bs x_0$ is called an \textit{attracting fixed point} of $T$ if there is a neighborhood $U$ of $\bs x_0$ such that for any $\bs x\in U$, the orbit of $\bs x$ under $T$ is contained in $U$ and converges to $\bs x_0$. The \textit{basin of attraction} of $\bs x_0$ under $T$, denoted $U_T(\bs x_0)$ is the largest such neighborhood $U$. Let $T_{K,\infty}$ denote the operator defined by $\lim_{n\to\infty}T_{K,n}(\bs x)$, for all $\bs x$. $T_{K,\infty}(\bs x)$ represents the estimated value obtained from an infinite time series after imputing the missing data using parameter $\bs x$. Observe that the strong consistency of $\hbs\gamma$ implies that $\bs\gamma_0$ is the only fixed point of $T_{K,\infty}$, that is $T_{K,\infty}(\bs \gamma_0)=\bs \gamma_0$, and $T_{K,\infty}(\bs x)\neq\bs\gamma_0$, for $\bs x\neq\bs\gamma_0$, for all $K$.  In this case, the infinite sample is obtained by imputing using the true $\bs\gamma_0$, the exact past values ($\F_0$) yielding a correctly specified reconstructed time series.

From what we discussed so far, the role that $s$ and $n$ play in the convergence of Algorithm 1 become clear. The role of $s$ is to guarantee that, starting at $\hbs\gamma_0$ in some appropriate set and under suitable conditions, as the algorithm is iterated the estimated values $\hbs\gamma_{s,n}$ will travel through $\Omega$, eventually arriving at $\bs \gamma_0$. The limit in $n$ along with the strong consistency of $\hbs\gamma$  guarantee that, when the orbit of $\hbs\gamma_0$ arrive at $\bs\gamma_0$, the estimation method will recognize so and stop the algorithm.

Notably, $K$ plays no role in deriving large sample results. In practice, however, $K$ speeds up the convergence of the algorithm by reducing the variance of the estimated values within each step of the algorithm. To see why this is the case, let $\F_{\mathcal I}=\sigma\{Y_i,i\in\mathcal I\}$ and observe that, given $\F_{\mathcal I}$, $\hbs\gamma_{s,n}^{1}, \cdots,\hbs\gamma_{s,n}^{K}$ are independent and identically distributed, since they share the same generating process and each imputation is independent from the previous. Under classical conditions, such as $\E_{\pi}\big(|\hbs\gamma|\big|\F_{\mathcal I}\big)<\infty$, where the expectation of a random vector is to be understood as the component-wise expectation, and $\E_{\pi}(\cdot)$ denotes the expectation taken in terms of the distribution of $\{Y_i, i\in \mathcal I\}$, the sequence satisfy the strong law of large numbers as $K\to\infty$. Indeed, observe that
\begin{equation*}
\var\big(T_{K,n}(\hbs\gamma_{s,n})|\F_\mathcal I\big)=\frac1K\var\big(\hbs\gamma_{s,n}^1|\F_\mathcal I\big) \quad \text{and}\quad \E_\pi\big(T_k(\hbs\gamma_{s,n})|\F_\mathcal I\big)=\E_{\pi}(\hbs\gamma_{s,n}^1|\F_\mathcal I),
\end{equation*}
for all $K>0$, with $\E\big(T_{K,n}(\hbs\gamma_{s,n})\big)=\E\big(\E_\pi(\hbs\gamma_{s,n}^1|\F_\mathcal I)\big)=\E(\hbs\gamma)$. Hence
\[\lim_{K\to\infty}T_{K,n}(\hbs\gamma_{s,n})=\E_\pi(\hbs\gamma_{s,n}^1|\F_\mathcal I),\]
almost surely. These observations were the main motivators for the development of the stopping criteria presented in Section \ref{sc}, which are based on the reduction in the variance between successive iterations.

For $\bs\gamma_0\in\Omega$ and $s,N\geq1$, let
\[U_{s,N}(\bs\gamma_0) := \bigg\{\bs x\in\Omega:\bigg[\bigcup_{i=N}^\infty\bigcup_{j=s}^\infty\big\{T_{K,i}^j(\bs x)\big\}\bigg]\cap U_{T_{K,\infty}}(\bs\gamma_0)\neq\emptyset\bigg\}.\]
We make the following assumptions:
\begin{itemize}
\item[\textbf{A1.}] There exists $s_0>0$ such that $T_{K,n}^s\to T_{K,\infty}^s$ uniformly in some compact neighborhood of $\bs\gamma_0$, for all $s>s_0$.
\item[\textbf{A2.}] The initial point $\hbs\gamma_0$ is chosen such that $\hbs\gamma_0\in U_{s_1,n_1}(\bs\gamma_0)$, for some $s_1>0$ and $n_1>0$.
\end{itemize}
Assumptions A1 and A2 are high level ones intended to maintain the discussion as general as possible without having to rely in a specific subclass of models, or a single choice of the estimator $\hbs\gamma$ to be applied. Assumption A1 is a very mild assumption on the behavior of $\hbs\gamma$ in the vicinitudes of $\bs\gamma_0$. It guarantees that $\lim_{n\to\infty} T_{K,n}^s(\bs x)=T_{K,\infty}^s(\bs x)$ for all $\bs x$ in a neighborhood of $\bs\gamma_0$, a crucial step in proving the convergence of Algorithm 1.

Assumption A2 defines the set of ``good'' initial points alluded in the beginning of this section. If  $\bs x\in U_{s_1,N_1}(\bs\gamma_0)$, it means that, after a sufficiently large number of iterations, i.e. $s>s_1$, considering a sufficiently large sample size $n>N_1$, $T_{K,n}^s(\bs x)$ will enter the orbit of $\bs\gamma_0$ under $T_{K,\infty}$ never to leave. This is to counter a rather pathological situation in which the basin of attraction of $T_{K,n}$ could start shrinking and migrating away from $U_{T_{K,\infty}}(\bs\gamma_0)$, in which case $\hbs\gamma_0\in U_{T_{K,\infty}}(\bs\gamma_0)$ but $\lim_{n\to\infty}\lim_{s\to\infty}T_{K,n}^s(\hbs\gamma_0)\neq\bs\gamma_0$.

The next theorem summarizes the convergence of Algorithm 1. We implicit assume the context and notation developed in this section.
\begin{thm}
Under conditions A1 and A2, Algorithm 1 converges almost surely to $\bs\gamma_0$ as $n$, and $s$ goes to infinity, for any $K>0$.
\end{thm}
\noindent\textbf{Proof.} We will show that, for suitable initial points $\hbs\gamma_0$, $\lim_{n\to\infty}\lim_{s\to\infty}T_{K,n}^s(\hbs\gamma_0)$ and $\lim_{s\to\infty}\lim_{n\to\infty}T_{K,n}^s(\hbs\gamma_0)$  both exist and are equal to $\bs\gamma_0$ with probability 1, for all $K>0$. We start by showing that $\lim_{s\to\infty}\lim_{n\to\infty}T_{K,n}^s(\hbs\gamma_0)=\bs\gamma_0$.  Let $K>0$ be fixed and suppose we have an infinite sample. The strong consistency of $\hbs\gamma$ implies that $\bs\gamma_0$ is the only fixed point of $T_{K,\infty}$. Let $s^\ast:=\max\{s_0,s_1\}$, and take $\hbs\gamma_0\in U_{s^\ast,n_1}(\bs\gamma_0)$ as in A2. A1 imply that  $\lim_{n\to\infty} T_{K,n}(\bs x)=T_{K,\infty}(\bs x)$ almost surely, for all $\bs x$ in a compact neighborhood of $\bs\gamma_0$, whereas A2 implies that $\hbs\gamma_{s,n}\in U_{T_{K,\infty}}(\bs\gamma_0)$, for all $s>s^\ast$. Hence
\begin{equation}\label{bas}
\lim_{s\to\infty}\lim_{n\to\infty} T_{K,n}(\hbs\gamma_{s,n})=\lim_{s\to\infty} T_{K,\infty}^s(\hbs\gamma_{0}) = \bs\gamma_0, \quad \text{almost surely}.
\end{equation}
For the other limit, we invoke A2 which allows the interchange of the limits in \eqref{bas}, yielding $\lim_{n\to\infty}\lim_{s\to\infty}T_{K,n}^s(\hbs\gamma_0)=\bs\gamma_0$ almost surely, completing the proof. \hfill\qed

Before we proceed to discuss some practical aspects regarding Algorithm 1, some remarks are in order.
\begin{rmk}
In Step 5, we average $\hbs\gamma_{s,n}^{1}, \cdots,\hbs\gamma_{s,n}^{K}$  to define $\hbs\gamma_{s+1,n}$. In principle, the transformation $T_{K,n}$ that takes $\hbs\gamma_{s,n}$ and returns $\hbs\gamma_{s+1,n}$ could be substituted by any continuous transformation, say $\Lambda:\Omega\to\Omega$, defined via $\hbs\gamma_{s+1,n}=T_{K,n}(\hbs\gamma_{s,n} )=\Lambda(\hbs\gamma_s^{1}, \cdots,\hbs\gamma_s^{K})$. A similar result as in Theorem 1 can be obtained, assuming for instance, that $\E\big(\big|\Lambda(\hbs\gamma_s^{1}, \cdots,\hbs\gamma_s^{K})_i\big|\big)<\infty$,  for $i\in\{1,\cdots,d\}$.
\end{rmk}
\begin{rmk}
In practice, the main effects of small samples is related to obtaining the initial guess $\hbs\gamma_0$. In finite samples, the smaller the $n$ and the greater the $m$, the harder is to obtain a ``good'' estimate $\hbs\gamma_0$. Also, the higher the proportion of missing data, the more iterations will be needed to improve the estimation from $\hbs\gamma_0$ to the point of convergence. The missing data mechanism do not interfere with the results in Theorem 1. But in practice it can add another layer of difficulty - see Section \ref{sv}, where we discuss this point in more details.
\end{rmk}
\begin{rmk}
Observe that, for any $\bs a, \bs a_1,\cdots,\bs a_{s-1}\in\Omega$, and $A\subset\Omega$
\begin{align*}
P\big(T_{K,n}^{s+1}&(\hbs\gamma_0)\in A|T_{K,n}^s(\hbs\gamma_0)=\bs a,\cap_{i=1}^{s-1}\big[T_{K,n}^s(\hbs\gamma_0)=\bs a_i], \F_{\mathcal I} \big)\\
&=P\big(T_{K,n}(\bs a)\in A|T_{K,n}^s(\hbs\gamma_0)=\bs a,\cap_{i=1}^{s-1}\big[T_{K,n}^s(\hbs\gamma_0)=\bs a_i], \F_{\mathcal I} \big)\\
&= P\big(T_{K,n}(\bs a)\in A|T_{K,n}^s(\hbs\gamma_0)=\bs a, \F_{\mathcal I} \big) =P\big(T_{K,n}^{s+1}(\hbs \gamma_0)\in A|T_{K,n}^s(\hbs\gamma_0)=\bs a, \F_{\mathcal I} \big).
\end{align*}
This shows that, conditionally on $\F_{\mathcal I}$, $\{T_{K,n}^s\}_{s=0}^\infty$ is a Markov chain with general state space $\Omega\subseteq\R^d$. Hence, Theorem 1 could be studied using tools from Markov chain theory  \citep[see, for instance,][]{meyn}. However, we adopted a more dynamical-system-oriented approach because it requires fewer assumptions.
\end{rmk}
\section{Practical Considerations}
The proposed methodology depends on some prespecified quantities, namely, the starting value $\hbs\gamma_0$; $K$, the number of multiple imputations to be performed; a stopping criteria and $H$, the maximum number of iterations allowed. In what follows we discuss some strategies to determine these values in practice. In this section we adopt the notation used in the presentation of Algorithm 1.
\subsection{Starting value}\label{sv}
The first step in applying Algorithm 1 is to provide the initial estimate of $\bs\gamma_0$, $\hbs\gamma_0$. There are several ways to obtain this estimate in practice. One simple solution is to use the estimate obtained upon applying the estimation procedure to the longest sequence of contiguous observed values in the sample.

More formally, suppose that $\mathcal{I}$ can be partitioned into $U$ sub-intervals of consecutive indices, say $\mathcal{I}_1,\cdots,\mathcal{I}_U$, with $\mathcal{I}_u = \{t_{u,1},\cdots,t_{u,v_u}\}$ denoting the $u$-th interval of size $v_u$. Define
\begin{equation*}
    u^\ast := \underset{1\leq u \leq U}{\mathrm{argmax}}\{v_u\}.
\end{equation*}
When $u^\ast$ is unique, $\mathcal{T}:=\mathcal{I}_{u^*}$ is the set of indices corresponding to the longest sequence of non-missing data in the time series. We propose estimating $\hbs\gamma_0$ using the subset $\{Y_t\}_{t \in \mathcal{T}}$. In cases where two or more sequences have the same maximum length, one can either arbitrarily select one of them or average the estimates obtained from each maximum-length subset to define $\hbs\gamma_0$.

In some cases, obtaining $\hbs\gamma_0$ from the longest sequence of contiguous observed values may result in a poor initial value for the algorithm, particularly when the sequence is too short - which is usually the case when $N$ is small and/or $m$ is large comparatively. This can also be influenced by the missing data mechanism. For instance, if the missing data mechanism impose that every $p$ time steps, there is a missing data, the longest sequence of contiguous points is of size $p$. On the other hand, if the missing data mechanism is MCAR and we are using a consistent estimator, then the longest sequence of contiguous observed values is itself consistent, as in this case, as $\lim_{n\to\infty}\#\mathcal{T} = \infty$.

In such scenarios, an alternative approach is necessary. One viable solution is to consider the longest sequence obtainable from the original time series by imputing at most $L$ consecutive points, where $L$ is small (e.g., 1 or 2), using any automatic imputation method. This approach is valid as long as the number of imputed values is small relative to the total length of the time series.

\subsection{Number of multiple imputations}

The next step in the implementation of the proposed method is to determine a value for the number of multiple imputations ($K$) to be used in Step~\ref{it:4}. After $m$ iterations of Steps~\ref{it:2} to~\ref{it:5}, the estimates $\hbs\gamma_{m-1}^1,\cdots,\hbs\gamma_{m-1}^K$ are obtained and averaged to define $\hbs\gamma_m$. Therefore, $K$ should be sufficiently large to ensure that the average of the $K$ estimates converges to a stable value, which will then be used as the updated parameter value for the next iteration of Steps~\ref{it:2} to~\ref{it:5}.

Note that the sequence $\hbs\gamma_{m-1}^1,\cdots,\hbs\gamma_{m-1}^K$ may not be independent, as these estimates are computed from time series that differ only in the set $\mathcal{I}^c$. In the context of GARMA models presented in the simulation, the value of $K$ was determined through a pilot simulation study. Even when the proportion of missing values was as high as 70\%, we found that $K=25$ provided a good balance between computational efficiency, while ensuring the convergence of the sample mean to a stable value. Additionally, our simulation revealed evidence of short-range dependence in the sequence $\hbs\gamma_{m-1}^1,\cdots,\hbs\gamma_{m-1}^K$. While a much larger value of $K$ could also be used, especially in practical applications where there is no need for thousands of replications, $K=25$ proved to be sufficient for our purposes.

\subsection{Stopping Criteria}\label{sc}

We propose two convergence criteria, as described in the following subsections. Both criteria consider the reduction in variability of the estimated parameter as the algorithm iterates. Low variability in the estimated parameter is naturally associated with the convergence of the iterative process (see Section \eqref{TD}). The question we aim to answer is which criterion is more efficient in detecting process convergence and what the difference is in terms of the estimated value once both criteria have been satisfied.

\subsubsection{Coefficient of Variation} 

The coefficient of variation-based stopping criterion (CVSC) relies on the coefficient of variation of the Euclidean distance between the estimate vectors. At each iteration of the algorithm, after computing the component-wise mean of the estimate vectors $\hbs\gamma_{h-1}^1,\cdots,\hbs\gamma_{h-1}^K$, the Euclidean distance is calculated between the estimated vector of the current iteration and that of the immediately preceding iteration. The coefficient of variation of these distances is computed at each iteration, the stopping criterion is considered met once the relative distances fall below a pre-specified tolerance, in which case the estimate vector of that iteration is taken as the final estimate associated with that multiple imputation. A pseudo-algorithm for CVSC is as follows:

\begin{algorithm}[H]
\vspace{0.3cm}
\begin{enumerate}[leftmargin = 0.5em]
\item Obtain the initial estimate vector $\hbs\gamma_0$;
\item Compute $\hbs\gamma_1$ following steps \ref{it:2} to \ref{it:5} of Algorithm 1;
\item Calculate $d_1=d(\hbs\gamma_0,\hbs\gamma_1)$, where $d$ denotes the Euclidean distance;
\item Compute $\hbs\gamma_2$ by iterating steps \ref{it:2} to \ref{it:5} of Algorithm 1;
\item Calculate $d_2=d(\hbs\gamma_1,\hbs\gamma_2)$ and $C_1$, the coefficient of variation of the set $\{d_1,d_2\}$, that is,
\begin{equation*}
    m_1 = \frac12 \sum_{i=1}^2d_i,\quad S_1^2=\sum_{i=1}^2(d_i-m_1)^2, \quad C_1=\frac{S_1}{m_1};
\end{equation*}
\item Iterate the algorithm until $|C_h/C_{h-1}-1|<\tau$, for some tolerance $\tau\in(0,1)$.
\end{enumerate}
\caption{Coefficient of Variation-based stopping criteria - CVSC}
\end{algorithm}
The tolerance value $\tau$ in step 6 represents the relative rate of decline in the variability of the Euclidean distances between successive parameter estimates and controls how close consecutive coefficient of variations must be in terms of proportion in order to declare convergence of the algorithm.

\subsubsection{Variance reduction stopping criterion}
The second criterion we consider is based on the reduction of variance in consecutive estimates (VRSC). The idea is that as the algorithm iterates, we track the behavior of the variance vector associated with the estimated parameters at each iteration. Specifically, at each step, we compute the Euclidean distance between the current variance vector and that of the previous iteration. The algorithm is considered to have met the stopping criterion when the absolute difference between consecutive distances falls below a predefined tolerance. The VRSC stopping criterion can be systematized as follows:

\begin{algorithm}[H]
\vspace{.3cm}
\begin{enumerate}[leftmargin = 0.5em]
    \item Obtain the initial estimate vector $\hbs\gamma_0$;
    \item Obtain $\hbs\gamma_1$ by following steps \ref{it:2} to \ref{it:5} of Algorithm 1;
    \item Calculate the vector $\bs S_1^2$ containing the component-wise sample variance of the set $\{\hbs\gamma_0,\hbs\gamma_1\}$;
    \item Obtain $\hbs\gamma_2$ by iterating steps \ref{it:2} to \ref{it:5} of Algorithm 1;
    \item Calculate $\bs S_2^2$, the vector containing the component-wise sample variances of the set $\{\hbs\gamma_0,\hbs\gamma_1,\hbs\gamma_2\}$. Compute $d_1 = d(\bs S_1^2, \bs S_2^2)$, i.e., the Euclidean distance between the vectors $\bs S_1^2$ and $\bs S_2^2$;
    \item Iterate the algorithm until $|d_h - d_{h-1}| < \tau$, for some tolerance $\tau>0$.
\end{enumerate}
\caption{Variance reduction stopping criteria - VRSC}
\end{algorithm}
The tolerance $\tau$ in the VRSC has a different meaning to the CVSC as it  considers the difference between consecutive euclidean distances calculated from consecutive variance vectors. As such, the CVSC tends to be more conservative regarding convergence, often requiring more iterations than VRSC for the same threshold. A comparison between these two criteria in the context of GARMA model is presented in Section \ref{sec:ar}.

\subsection{Handling Non-Convergence and the Maximum Number of Iterations}

Algorithm 1 may not always meet the convergence criterion within the predefined maximum number of iterations. In such cases, simple diagnostic measures can be employed to assess and address the issue. A practical approach is to restart the algorithm using the last estimated value as the initial state -- this is equivalent to increasing $H$ from the start but is computationally more efficient.

Another factor that may affect convergence is the choice of the initial state, particularly in time series with a high percentage of missing values, where obtaining a reliable initial parameter estimate can be challenging. In these cases, restarting the algorithm using the last obtained parameter may improve convergence. Additionally, the stopping criteria can be reviewed, either by adjusting the threshold or modifying the criterion itself.

For likelihood-based estimation procedures, the algorithm may fail to converge if it becomes trapped in a region of the likelihood surface with little variation, preventing updates from triggering the stopping criteria. This can be diagnosed by plotting the sequences obtained in step \ref{it:4}. If necessary, increasing the value of $K$ or modifying the initial state might help.

In our simulations, we found that setting $H = 30$ iterations was generally sufficient to achieve high convergence rates across all scenarios with the applied thresholds. As expected, the proportion of non-convergent cases increased with the percentage of missing data. However, even at 70\% missing values, the highest observed non-convergence rate was 2.3\% for the CVSC and 3.7\% for the VRSC, which we consider acceptably low. For further details, please refer to Section \ref{iter}.

\section{Monte Carlo Simulation Study}
In this section we present a simulation study in the context of GARMA models. To the best of our knowledge, this is the first study on GARMA models with missing data in the literature. We will consider 3 GARMA-like models  with systematic components given by \eqref{arma} with errors in the predictor scale, meaning $r_t:=g(y_t)-g(\mu_t)$, with $g$ being the logit function. In the following we present the models from the oldest to newest. The $\beta$ARMA (Beta ARMA) model was introduced in \cite{Rocha2009}. For the $\beta$ARMA, the Beta distribution is parametrized in terms of its conditional mean, denoted by $\mu_t\in(0,1)$, and a shape parameter $\nu>0$, resulting in  the density
\begin{equation*}
f(y;\mu_t,\nu|\F_{t-1})=\frac{\Gamma(\nu)}{\Gamma(\nu\mu_t)\Gamma\big(\nu(1-\mu_t)\big)}\,y^{\nu\mu_t-1}(1-y)^{\nu(1-\mu_t)-1}I(0<y<1).
\end{equation*}
The KARMA (Kumaraswamy ARMA) model was introduced in \cite{Bayers}. For the KARMA, the Kumaraswamy distribution is parametrized in terms of its conditional median, represented by $\mu_t\in(0,1)$, and a shape parameter $\nu>0$. The conditional density is given by
\begin{equation*}
f(y;\mu_t,\nu| \F _{t-1}) =\frac{\varphi\log (0.5)}{\log(1-{\mu}_t^\nu)} {y}^{\nu-1} (1-{y}^\nu)^{\frac{\log (0.5)}{\log(1-{\mu}_t^\nu)} -1}I(0<y<1).
\end{equation*}
The UWARMA (Unit-Weibull ARMA) model was introduced in \cite{uw}. For the UWARMA, the Unit-Weibull distribution is parameterized in terms of its $\rho$th quantile, represented by $\mu_t\in(0,1)$, and a shape parameter $\nu>0$, for any fixed $\rho\in(0,1)$. The conditional density on a UWARMA model is given by
\begin{equation*}
f_\rho(y;\mu_t,\nu|\F_{t-1})=\frac\nu y\bigg[\frac{\log(\rho)}{\log(\mu_t)}\bigg]\bigg[\frac{\log(y)}{\log(\mu_t)}\bigg]^{\nu-1}\rho^{\big[\frac{\log(y_t)}{\log(\mu_t)}\big]^\nu}\!\!I(0<y<1).
\end{equation*}
The $\beta$ARMA, KARMA and UWARMA were chosen because these models are featured in the package \texttt{BTSR} \citep{BTSR} for R \citep{r}, which makes them freely available for anyone to use. Although $\beta$ARC models are also available in \texttt{BTSR}, they were not included in the simulation as they present systematic component whose behavior is radically different from \eqref{arma}, making comparison impossible with the other models.

It is worth mentioning that inference in KARMA models considered the more restrictive approach of conditional maximum likelihood (CMLE). Originally in \cite{Rocha2009}, the authors also considered CMLE, but the $\beta$ARMA models are a particular case of the $\beta$ARFIMA models of \cite{PUMI2019}, where the authors considered inference via partial maximum likelihood (PMLE), adopted here. Computationally speaking, the difference between using PMLE or CMLE is immaterial.

\subsection{Data Generating Process}

Using the \texttt{BTSR} package \citep[][version 1.0.0]{BTSR}, time series of size $n=500$ were generated from models $\beta$ARMA, KARMA, and UWARMA models, with order $p=q=1$, considering a burn-in of size 100. Three scenarios were established for the simulations, each corresponding to a parameter vector $\bs\gamma = (\alpha, \phi, \theta, \nu)'$ with $\bs{\gamma}_1 = (0.5, -0.4, -0.6, 20)'$, $\bs{\gamma}_2 = (0.5, -0.5, -0.3, 30)'$, and $\bs{\gamma}_3 = (0.5, -0.4, -0.2, 20)'$. For simplicity, the same parameter $\nu$ was used to denote the fixed parameter of the three models. Missing data were introduced at proportions of $r \in \{0.1, 0.4, 0.7\}$, assuming that they occur completely at random. For simplicity, we assume that the first and last values of the time series were never missing.

These time series with missing data were then subjected to the multiple imputation process described in Algorithm 1, resulting in imputed series, on which parameter estimation using PMLE was performed. The data were generated using the stopping criteria described in Section \ref{sc}. In both cases, the tolerance was fixed at $\tau=0.01$ and the maximum number of iterations was set to $H=30$ and $K$ was set to 25 following the results of a pilot simulation (not shown). Within the \texttt{BTSR} package, optimization was conducted using Nelder-Mead. The simulation was carried out using the \texttt{R} language \citep{r}, version 4.4.1. For each combination of model, scenario, and proportion of missing data, the process was replicated $R=1,000$ times.
\section{Simulation Results}\label{sec:ar}
The simulation results are presented in Figures \ref{fig:sce1_barma} below and Figures 1 and 2 in the supplementary material. Each figure consists of twelve blocks. Each block represents a combination of parameter and proportion of missing data, described respectively in the columns and rows. Each block contains three box plots of the estimated parameters, two of which correspond to the stopping criteria (CVSC and VRSC, discussed in Section \ref{sc}), and one referring to the estimates based on the complete series. It is worth noting that the latter remain the same across the rows and serve only as a reference.
\subsection{Scenario 1}
\begin{figure}[ht]
    \centering
    \includegraphics[width = \textwidth]{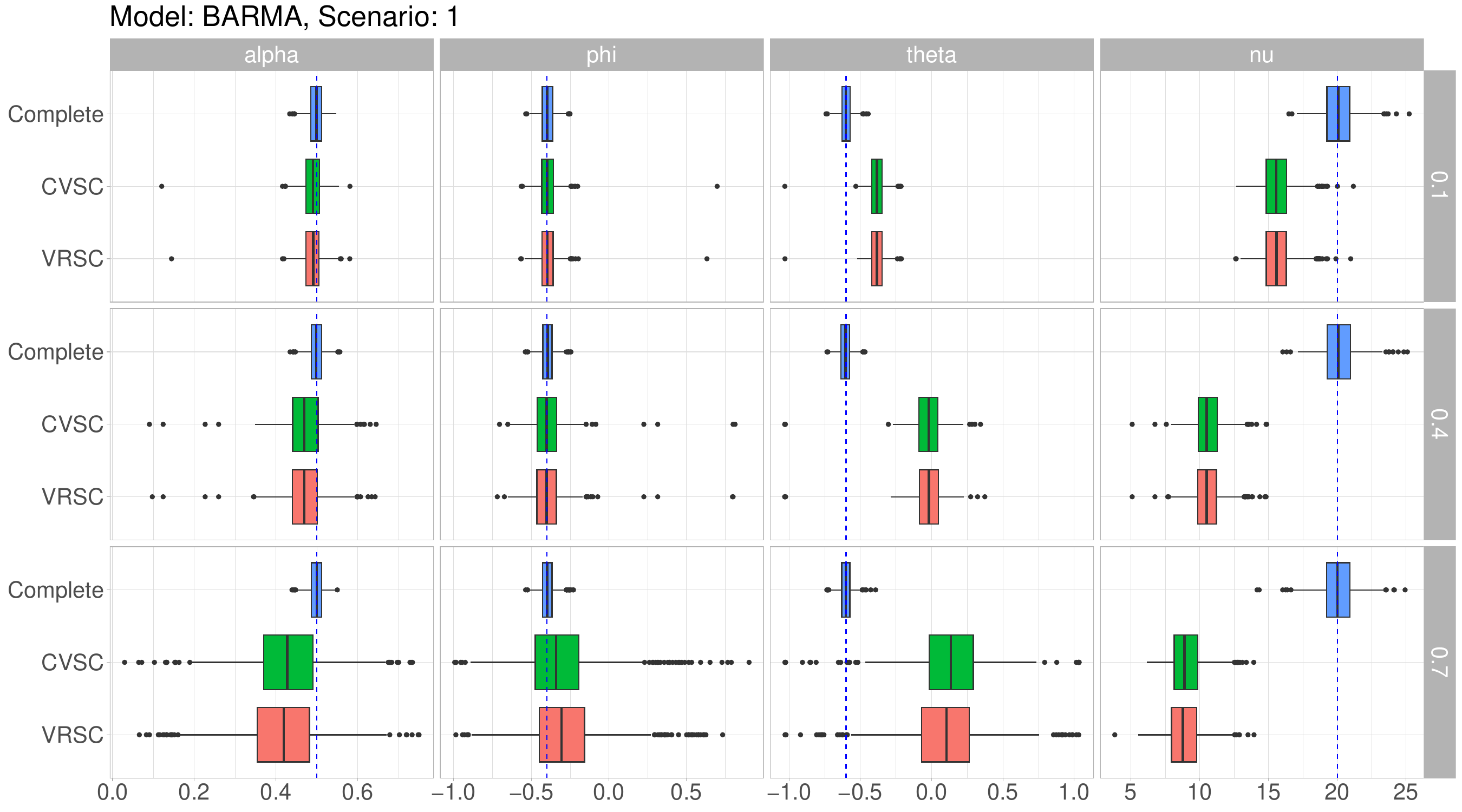}
    \caption{Simulation results for the $\beta$ARMA model under scenario 1.}
    \label{fig:sce1_barma}
\end{figure}

Simulation results are presented in Figures \ref{fig:sce1_barma} for the $\beta$ARMA and KARMA (displayed together due to similarity) and \ref{fig:sce1_uwarma} for the UWARMA. For the three models, as the proportion of missing values increase, we observe an increase in bias and variance across all cases, except for parameter $\phi$ in the context of $\beta$ARMA and $r\in\{0.1,0.4\}$, where the bias was not significantly affected by the missing values. Notably, the results for $\theta$ approached 0 as $r$ increased to 0.7.
\begin{figure}[ht]
    \centering
    \includegraphics[width = \textwidth]{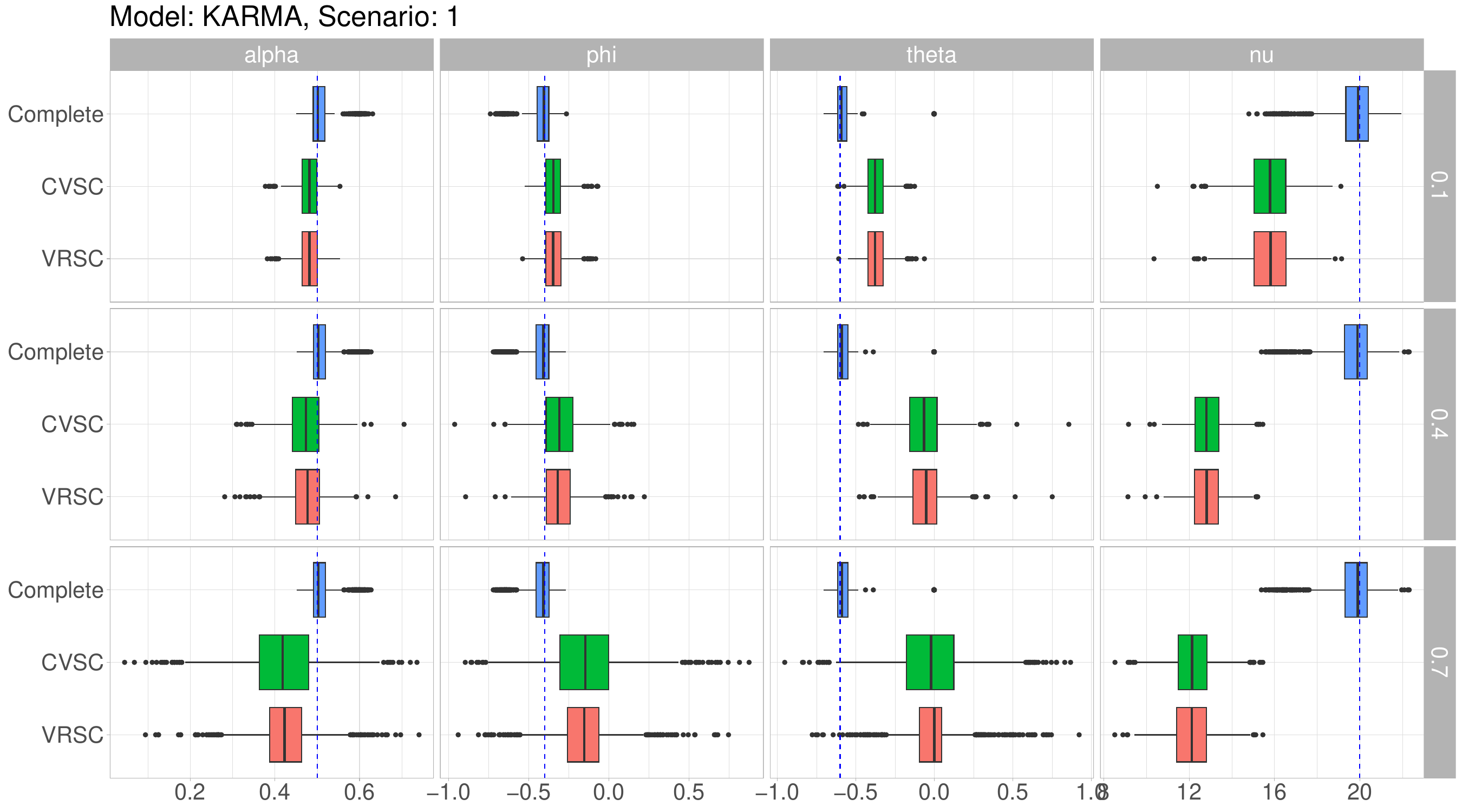}
    \caption{Simulation results for the KARMA model under scenario 1.}
    \label{fig:sce1_karma}
\end{figure}
\FloatBarrier
\begin{figure}[ht]
    \centering
    \includegraphics[width = \textwidth]{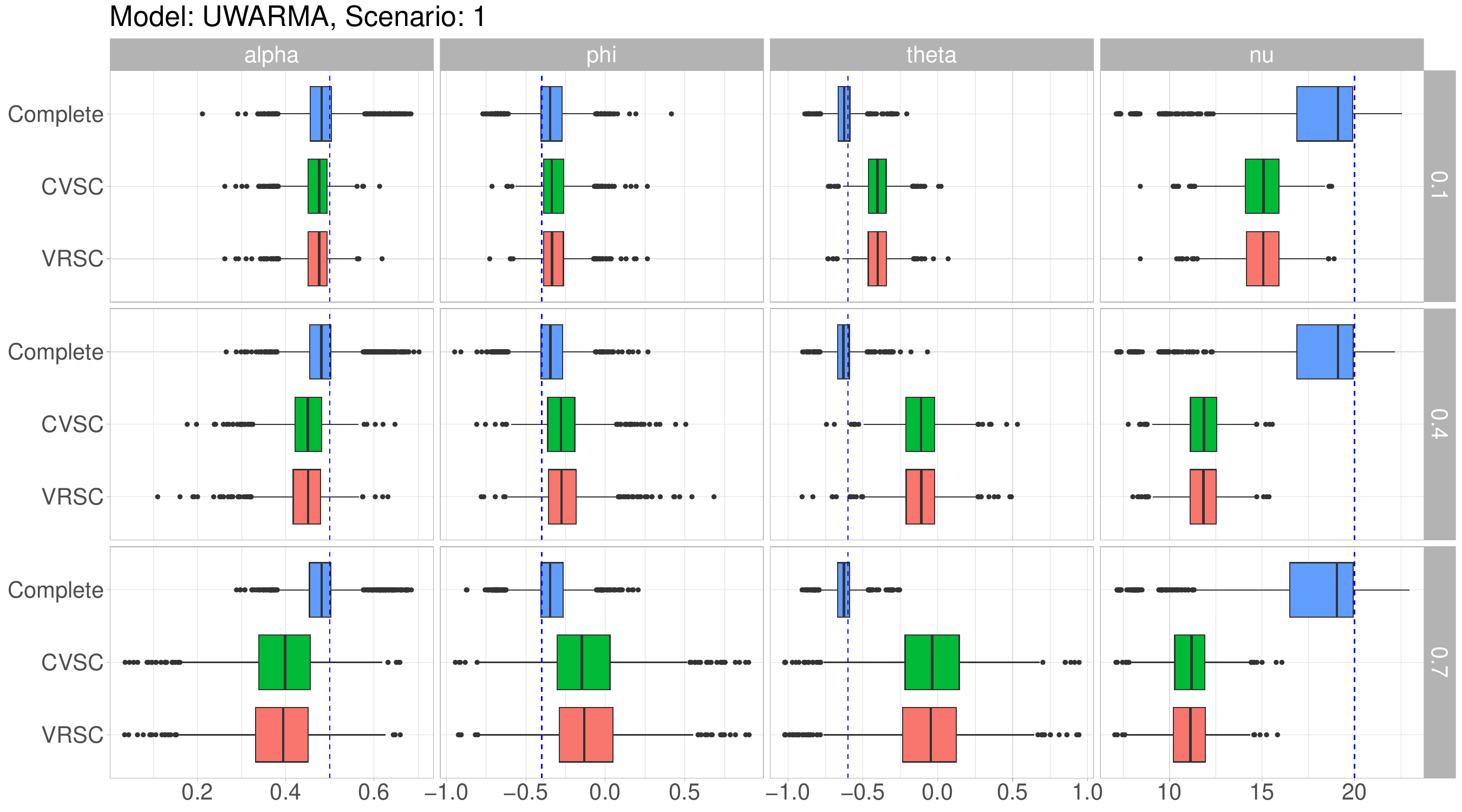}
    \caption{Simulation results for the UWARMA model under scenario 1.}
    \label{fig:sce1_uwarma}
\end{figure}

As for  the stopping criteria, for $r\in\{0.1,0.4\}$ both present similar behavior in terms of bias and variance for all models. For $r=0.7$, the results are mixed. Both stopping criteria presented quite similar behavior for the UWARMA, the CVSC presented slightly better results for the $\beta$ARMA whereas VRSC presented smaller variance and slightly smaller bias in the case of KARMA models.

\subsection{Summary of Scenarios 2 and 3}
Due to space limitations, we present here a brief summary of the main findings for Scenarios 2 and 3; complete results, including all figures, are available in the Supplementary Material.
\subsubsection*{Scenario 2.}
The results across all three models reveal several consistent patterns. First, bias and variance generally increase with the proportion of missing data ($r$), with $\nu$ consistently underestimated. The two stopping criteria (CVSC and VRSC) perform similarly for $r \in \{0.1, 0.4\}$, though their relative performance varies at $r=0.7$ depending on the model.

Notably, the KARMA model exhibits a distinctive behavior: for complete series, $\theta$ estimates remain fixed at the initial value of zero, suggesting a flat likelihood surface. However, introducing missing data at $r=0.10$ allows the optimization algorithm to move away from zero toward the true value, indicating that imputation sufficiently alters the likelihood surface to enable proper estimation. This phenomenon reverses at higher missingness rates ($r \geq 0.4$), where estimates return to zero.

\subsubsection*{Scenario 3.}
The patterns observed mirror those of Scenario 2, with bias and variance increasing in $r$. A key finding emerges for the KARMA model: at $r=0.10$, imputed series yield more accurate estimates for $\alpha$ and $\phi$ than complete series, despite increased variability. This is an undesirable, whilst common, finite sample effect in the estimation under the complete time series. For the UWARMA model, imputation improves $\theta$ estimation at low missingness ($r=0.10$), though this advantage dissipates as $r$ increases.

Across both scenarios, the stopping criteria produce nearly identical estimates in most cases, with VRSC showing a slight advantage only for the KARMA model at the highest missingness rate ($r=0.7$).
\subsection{Discussion}
The bias and variance of the estimates tended to increase as the proportion of missing data moves away from 10\%, which is expected due to the dominance of possibly inaccurate information due to imputation and the loss of information from the observed values. This effect is especially noticeable for $r=0.7$, whereas for $r=0.4$ often the effect was less noticeable for parameter $\alpha$ and $\phi$. Also noteworthy is the tendency of the PMLE to underestimate $\nu$, for all models and scenarios, with rare exceptions. For the UWARMA, both stopping criteria yielded nearly identical results, but overall both criteria performed similarly, especially when $r\in\{0.1,0.4\}$. For $r=0.7$, a marginal advantage for VRSC was observed.

An important facet of imputation in the context of time series analysis under massive amounts of missing data was also observed here. For all analyzed models, it was observed that as the percentage of missing data increases, the estimates of $\phi$ and $\theta$ tend to approach zero. This behavior is, to some extent, expected, since, although the proposed method takes into account the dependence structure reconstructed from the estimates obtained at each step to impute the series, the presence of a large volume of missing data -- as in the case of $r = 0.7$ -- ends up limiting the effectiveness of the imputation. In this context, imputation based on approximate parameter values is unable to adequately restore the correlation structure in $\mu_t$, resulting in an imputed series close to a uncorrelated sample. Consequently, the estimates of $\phi$ and $\theta$, which are responsible for this structure, tend to zero. This effect was, overall more evident for the parameter $\theta$. This is also a consequence of the loss of information that occurs under massive percentage of missing values, which ultimately affect every imputation method. Nevertheless, the results obtained can still be considered satisfactory, given the extreme nature of the problem when $r=0.7$.

\section{Joint Asymptotic Behavior}
To heuristically assess whether the asymptotic normality of the PMLE holds under  our proposed methodology, we examine pairwise scatter plots and marginal distributions of the estimates. Figures 7–9 in the Supplementary Material present these diagnostics for the $\beta$ARMA, KARMA, and UWARMA models under Scenario 1 ($\alpha=0.5$, $\phi=-0.4$, $\theta=-0.6$, $\nu=20$) using the VRSC stopping criterion.

Three main patterns emerge. First, the increase in bias and variance as the proportion of missing data increases is very clear. Second, the estimates of $\alpha$, $\phi$, and $\theta$ exhibit strong pairwise correlations, while $\nu$ shows much weaker correlation with the other parameters—despite not being orthogonal to them in the models considered. This effect is a common occurrence in models from the GARMA class and the behavior is maintained. Finally, interestingly the histogram behavior when $r=0.7$ is more symmetric and smooth, although more disperse, when compared to smaller values of $r$.

\subsection{Iterations to Convergence}\label{iter}

Table \ref{tab:nloops} presents the average number of iterations until convergence for the replications in which the maximum number of iterations, $H = 30$, was not reached. The table includes results for all combinations of model type, stopping criterion, percentage of missing data, and scenario. The percentage of replications that reached the maximum number of iterations is provided in parentheses. Notably, the average number of iterations until convergence is systematically lower for the VRSC stopping criterion, with the CVSC required over 40\% more iterations to attain convergence in most cases.

As for non-converge, this percentage is quite low overall peaking at 2.3\% for the CVSC and 3.7\% for the VRSC. For $r=0.1$ the CVSC present uniformly higher percentage of non-convergence compared to VRSC, peaking at 0.8\%. The situation is reversed when considering scenario 2, for $r\in\{0.4,0.7\}$. For other cases, no clear pattern was found, with CVSC and VRSC alternating their position. Interestingly, the average number of iterations for the CVSC criterion increases with the proportion of missing data, $r$. In contrats, the average number of iterations to convergence for for the VRSC criterion shows homogeneity across all values of $r$.

\begin{table}[!ht]
\caption{Average number of iterations for convergence and the percentage of replications that reached the maximum number of iterations (in parenthesis).}\label{tab:nloops}\vspace{.3cm}
\centering
\setlength{\tabcolsep}{3pt}
\renewcommand{\arraystretch}{1}
\small
\begin{tabular}{c|c|cc|cc|cc|cc|cc|cc}
    \hline
    \multirow{2}{*}{$r$}&\multirow{2}{*}{Scenario} & \multicolumn{4}{c|}{$\beta$ARMA}  & \multicolumn{4}{c|}{KARMA}  & \multicolumn{4}{c}{UWARMA}   \\
     \cline{3-14}
      &  & \multicolumn{2}{c}{CVSC} & \multicolumn{2}{c|}{VRSC} & \multicolumn{2}{c}{CVSC} & \multicolumn{2}{c|}{VRSC} & \multicolumn{2}{c}{CVSC} & \multicolumn{2}{c}{VRSC}  \\
     \hline
  \multirow{3}{*}{0.1}
     & 1 & 11.58 & (0.3) & 7.61 & (0.0) & 11.24 & (0.3) & 6.37 & (0.0) & 11.22 & (0.8) & 6.70 & (0.0) \\
     & 2 & 11.55 & (0.6) & 9.99 & (0.2) & 11.41 & (1.2) & 8.33 & (0.0) & 11.22 & (0.4) & 8.76 & (0.0) \\
     & 3 & 11.21 & (0.5) & 7.31 & (0.1) & 11.30 & (0.2) & 5.58 & (0.0) & 11.24 & (0.5) & 6.06 & (0.0)  \\
 \hline
    \multirow{3}{*}{0.4}
     & 1 & 12.71 & (0.8) & 6.62 & (0.3) & 12.23 & (0.3) & 6.06 & (0.1) & 12.63 & (0.6) & 6.71 & (0.2)  \\
     & 2 & 12.77 & (0.6) & 9.57 & (1.6) & 12.50 & (0.5) & 8.25 & (0.7) & 12.64 & (0.6) & 8.43 & (1.2)  \\
     & 3 & 12.60 & (0.6) & 7.95 & (0.8) & 12.14 & (0.4) & 6.00 & (0.0) & 12.63 & (0.6) & 6.70 & (0.2)  \\
 \hline
 \multirow{3}{*}{0.7}
     & 1 & 13.82 & (1.4) & 5.70 & (1.7) & 13.08 & (1.2) & 5.56 & (0.9) & 14.19 & (1.4) & 6.27 & (0.7)  \\
     & 2 & 13.75 & (2.3) & 9.11 & (3.7) & 13.71 & (1.7) & 7.60 & (3.1) & 14.61 & (2.0) & 7.95 & (2.7)  \\
     & 3 & 13.73 & (1.9) & 7.99 & (2.7) & 13.25 & (0.8) & 5.98 & (1.2) & 14.16 & (1.3) & 6.30 & (0.6)  \\
 \hline
\end{tabular}
\normalsize
\end{table}
\FloatBarrier

Focusing first on the marginal distributions, the histograms for parameter $\nu$ closely resemble a normal distribution but are centered at increasingly underestimated values as $r$ increases across all models. A similar pattern is observed for parameters $\alpha$ and $\phi$, with both bias (in absolute value) and variance increasing as $r$ grows. Parameter $\theta$ exhibits the worst behavior, with estimates frequently concentrating around zero.

The scatter plots for parameter pairs $(\alpha,\phi)$, $(\alpha,\theta)$, and $(\phi,\theta)$ indicate a strong correlation among these estimates across all models, except for $(\alpha,\phi)$ in the $\beta$ARMA case. For $r\in\{0.1,0.4\}$, the scatter plots generally display a well-defined bivariate normal pattern. Scatter plots involving parameter $\nu$ exhibit a recognizable bivariate normal pattern, with weaker correlations between $\nu$ and the other parameters.

\section{Empirical Analysis}
In order to showcase the usefulness of the proposed methodology, we conducted an empirical analysis on two real dataset.
\subsection{Stored Energy}\label{stored}
 We consider the same data used in \cite{scher2020}, referring to the proportion of hydroelectric energy stored in southern Brazil from January 2001 to April 2017, with a sample size $n=196$. Missing data were introduced at random, following a uniform distribution on the time stamps $2,\cdots,195$ (for simplicity the first and last observation were assumed never missing) at proportions of $r \in \{0.2,0.4,0.7\}$. The time series plot of the data is presented in Figure  \ref{fig:ts_comp}, while Figure \ref{fig:ts_miss} present the same time series after randomly removing 70\% of the data. Next, the multiple imputation procedure was applied to these artificially created time series, along with the estimation of the parameters for models $\beta$ARMA$(1,1)$ and KARMA$(1,1)$. We used $K=25$, $H=30$ and $\tau=0.01$, as in the simulation. This procedure was repeated 1,000 times.

\begin{figure}[ht]
\centering
\includegraphics[width=\textwidth]{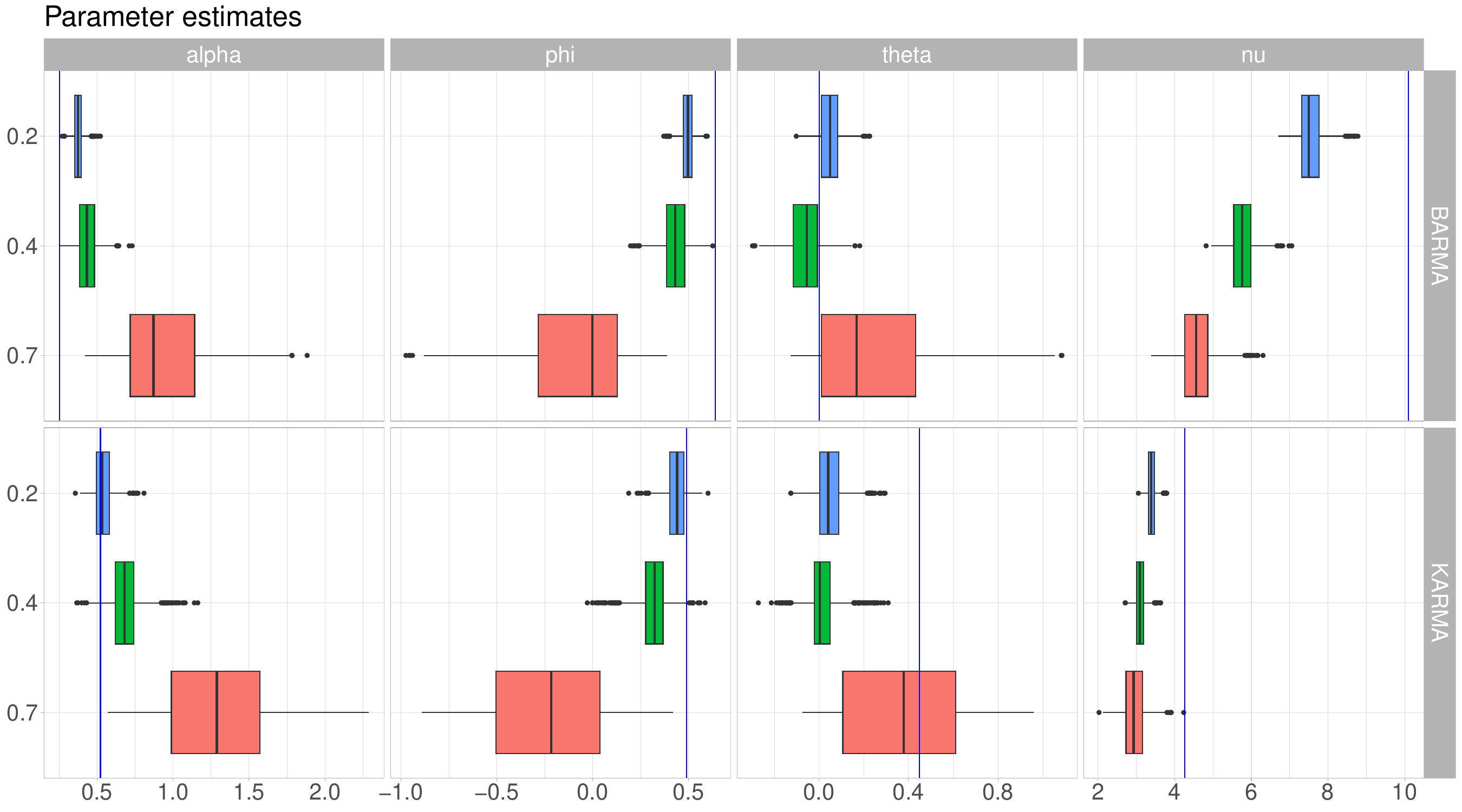}
\caption{Estimation results for the $\beta$ARMA and KARMA models, for proportion of missing data $r \in \{0.2,0.4,0.7\}$.} \label{fig:apl}
\end{figure}

\begin{figure}[ht]
    \centering
    \subfigure[Original time series ($n=196$).\label{fig:ts_comp}]{\includegraphics[width = \textwidth]{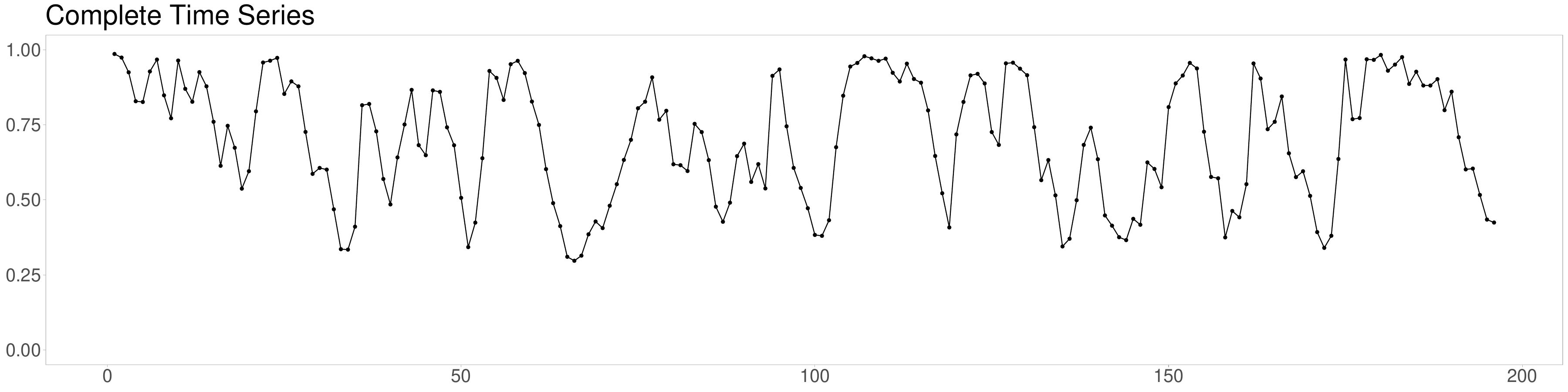}}\vspace{.3cm}
    \subfigure[Time series with 70\% of missing data.\label{fig:ts_miss}]{\includegraphics[width = \textwidth]{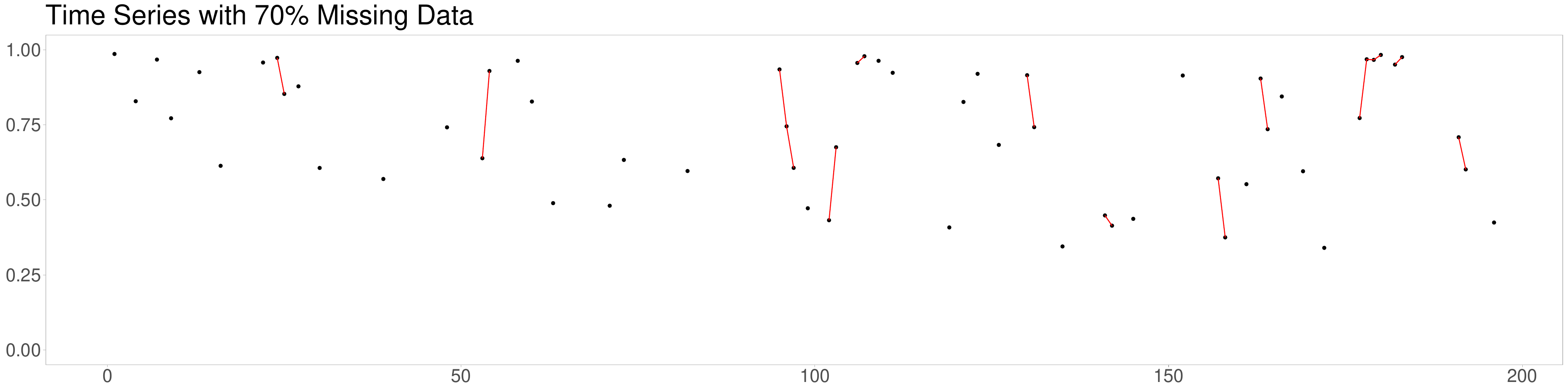}}
    \caption{Visual comparison between (a) the complete time series and (b) the same time series after randomly removing 70\% of the data.}
\end{figure}

The results of the empirical analysis are presented in Figure \ref{fig:apl}, which consists of eight blocks arranged in four columns corresponding to the parameters and two rows corresponding to the models. Each block contains three box plots, one for each proportion of missing data, along with a vertical line indicating the parameter estimate obtained from the complete series.

\FloatBarrier

For the $\beta$ARMA model, $\alpha$ presented higher estimated values when compared to the estimate obtained from the complete time series, with magnitude increasing with $r$. For parameters $\phi$ and $\nu$ the situation was reversed. The estimates of $\theta$ oscillate between the estimated value in the complete time series.

The KARMA model yielded estimates closer to the ones obtained from the complete time series for $\alpha$ and $\phi$ when $r=0.2$. However, introducing missing data at proportions $r\in\{0.4,0.7\}$ pulled the estimates for these parameters away from the estimates obtained from the complete time series. Regarding $\theta$, the estimates were concentrated around zero for $r\in\{0.2,0.4\}$, but for $r=0.7$, its median was closer to the estimate obtained from the complete series.

The estimates of $\nu$ in the KARMA model exhibited a pattern similar to that observed in the $\beta$ARMA model. A significant increase in the variability of the estimates for all parameters when $r=0.7$ was common to both models. For $r\in\{0.2,0.4\}$, the increase in variance was less noticeable, especially for the KARMA.

We emphasize that, in some cases, obtaining $\hbs\gamma_0$ from the longest sequence of contiguous observed values can negatively impact the initial value results, which plays an important role in the proposed methodology. This occurs when the value of $n$ is small and when the proportion of missing data $r$ is large, resulting in the longest sequence of observed data being very short, complicating estimation by the PMLE. Figures \ref{fig:ts_comp} and \ref{fig:ts_miss} illustrate this point. In this example, the longest sequences of contiguous values has only 4 points, a situation in which obtaining a good initial estimate is challenging.

\FloatBarrier

To overcome this problem, we can employ the solution presented in Section \ref{sv}. The idea is to consider the longest sequence of the original time series imputed with at most $L$ consecutive points, using any imputation method. To assess the impact of implementing this strategy, we conduct the same experiment as before, but considering this strategy  for $L \in \{0,1,2\}$, imputing values using the linear interpolation method implemented in package \texttt{imputeTS}. Observe that $L=0$ is equivalent to using the longest sequence without any modification presented in Figure \ref{fig:apl}.

\begin{figure}[ht]
    \centering
    \includegraphics[width = 0.9\textwidth]{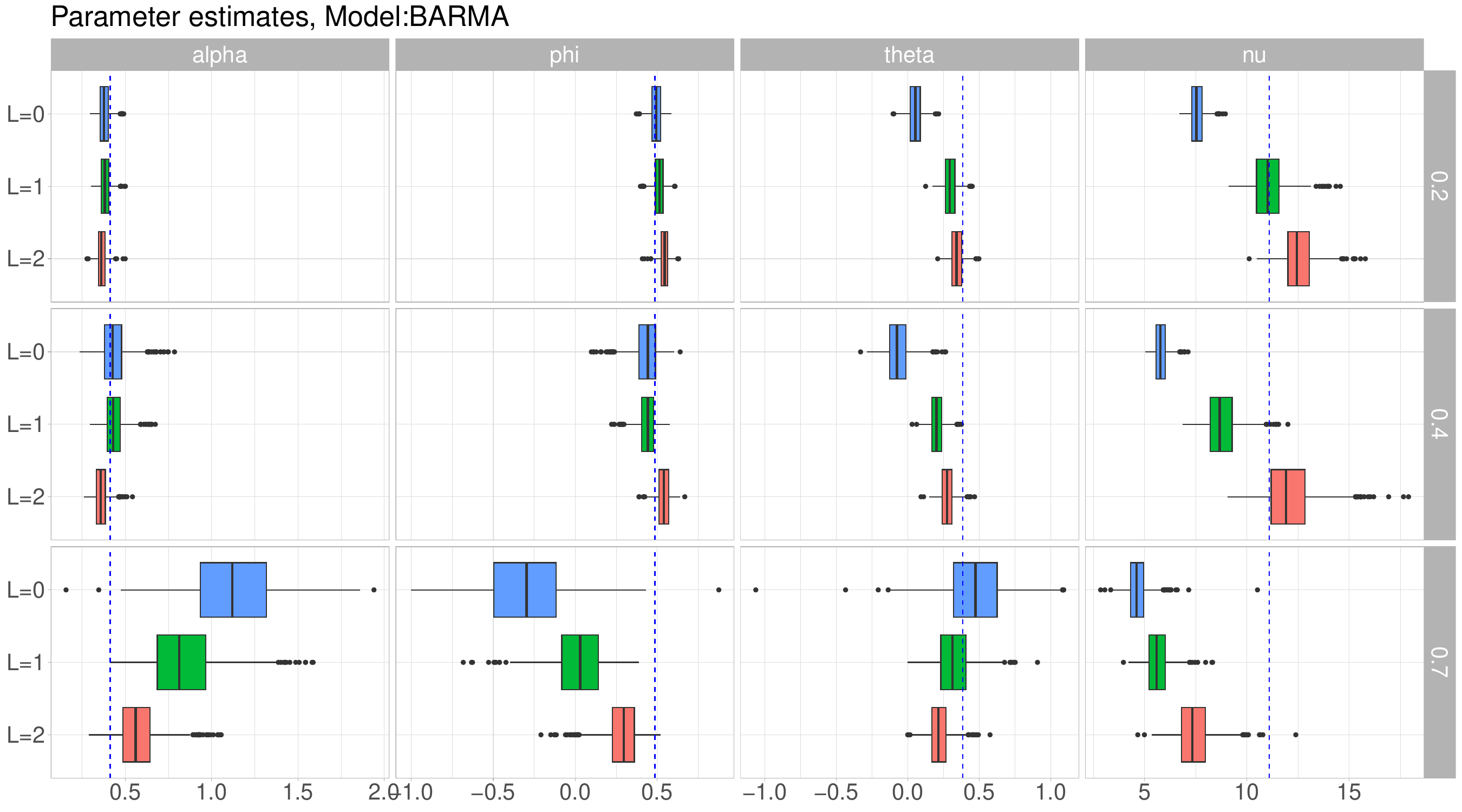}
    \caption{Experiment results considering the $\beta$ARMA model.}
    \label{fig:Lbarma}
\end{figure}

\begin{figure}[ht]
    \centering
    \includegraphics[width = 0.9\textwidth]{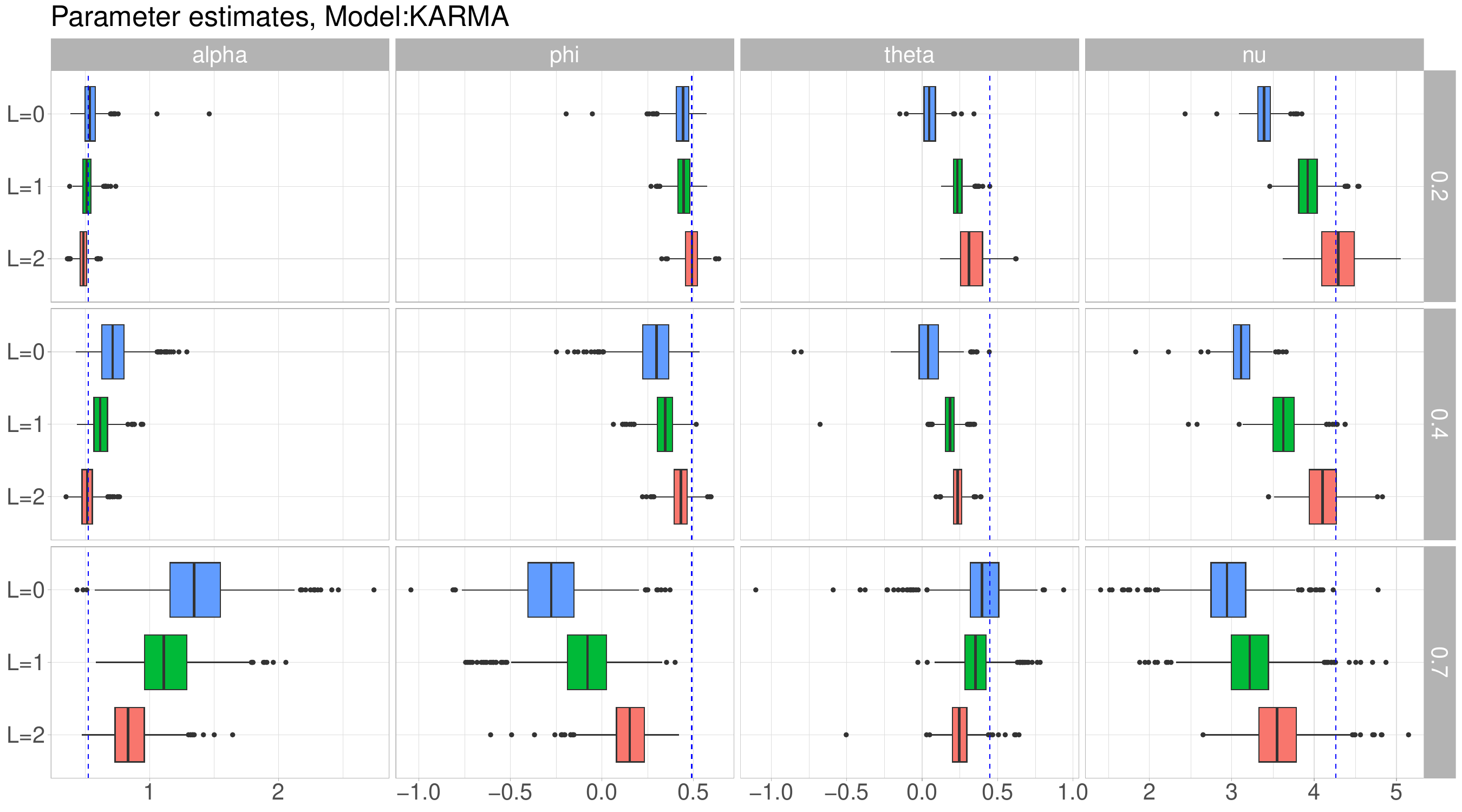}
    \caption{Experiment results considering the KARMA model.}
    \label{fig:Lkarma}
\end{figure}

The results are presented in Figures \ref{fig:Lbarma}, for the $\beta$ARMA, and \ref{fig:Lkarma} for the KARMA. Each figure contains twelve blocks, which are arranged in four columns representing the parameters and three rows representing the values of $r$. Each block contains three boxplots, each associated with one of the considered values of $L$. The vertical dashed blue lines represent the parameter estimates obtained from the complete series for the respective model, presented for reference.

\FloatBarrier

\FloatBarrier
For simplicity, we begin by analyzing the results for the KARMA model, presented in Figure \ref{fig:Lkarma}. For almost all parameters and all values of $r$, increasing the value of $L$ substantially improved the final estimated value, in the sense that it was closer to the values obtained from the complete time series. The only exception is $\theta$ when $r=0.7$, in which case increasing $L$ yielded worse results.

For the $\beta$ARMA model, the behavior is more complex. For the parameters $\alpha$ and $\phi$, when $r\in\{0.2,0.4\}$, using $L=0$ or $L=1$ yielded similar results, whereas $L=2$ resulted in slightly worse estimates. However, for $r=0.7$, increasing $L$ improved the estimates for both parameters.

For $\theta$, when $r\in\{0.2,0.4\}$, increasing $L$ improved the estimates. For $r=0.7$, $L=1$ yielded substantially better results than $L=0$, whereas for $L=2$, the results were worse than the previous ones.

For $\nu$, $L=0$ consistently produced the worst results. When $r=0.1$, $L=1$ yielded the best results, whereas for $r\in\{0.4,0.7\}$, $L=2$ provided the best estimates. Table \ref{nslice} presents the average length of the longest contiguous sequence for each value of $L$ used and $r \in \{0.2,0.4,0.7\}$. The average length is naturally a function of $r$. For $r=0.2$, the average length is 36, which is relatively low, though not extremely so. For $r=0.7$, however, this value drops to less than 12. In the most extreme case, $r=0.7$, the average length increases by approximately 50\% for each unitary increase in $L$, with this percentage increase being higher for lower values of $L$.
\begin{table}[!ht]
\setlength{\tabcolsep}{10pt}
\centering
\caption{Average length of the longest contiguous sequence.}\vspace{.3cm}
\label{nslice}
\begin{tabular}{c|ccc}
\hline
\backslashbox{$L$}{$r$} & $0.2$ & $0.4$ & $0.7$\\ \hline
$0$ & 36.0 & 22.3 & 11.9\\
$1$ & 84.9 & 42.4 & 17.6\\
$2$ & 149.0 & 73.8 & 25.8\\ \hline
\end{tabular}
\end{table}
\subsection{Air Pollution data}\label{pol}

In this section, we analyze time series data on the concentration of fine particulate matter of at most 2.5$\mu$m in the air, hereafter denoted as PM${}_{2.5}$. These fine particles are produced in construction sites, emitted by automobiles, roads, fields, smokestacks, fires, and other sources. In high concentrations, they can be harmful to human health. We consider data from a monitoring station located in Brandon, Manitoba, Canada, at coordinates 49°50'31.9"N, 99°55'08.0"W (station ID 70203). The data were obtained from the United States Environmental Protection Agency (EPA) AirNow program's ``developer tools'' website (https://docs.airnowapi.org/).

The dataset consists of daily averages (from midnight to midnight) of raw PM${}_{2.5}$ concentrations (in $\mu$g/m${}^3$) from January 1st, 2021 to November 11, 2023. The total sample size is 1045, of which only 527 data points are observed, while 518 are missing (49.57\%). The high proportion of missing data  makes this dataset particularly challenging and well-suited for demonstrating the imputation capabilities of our proposed algorithm. One observation (09-11-2023) was recorded as a negative value, which, according to AirNow's documentation, can occur under certain conditions due to ``instrument drift over time.'' We treat this as an instrumentation error and discard the observation, handling it as missing data. The observed concentrations range from 0.6 to 46.2$\mu$g/m${}^3$.

\begin{figure}[ht]
    \centering
    \mbox{
    \includegraphics[width = 0.8\textwidth]{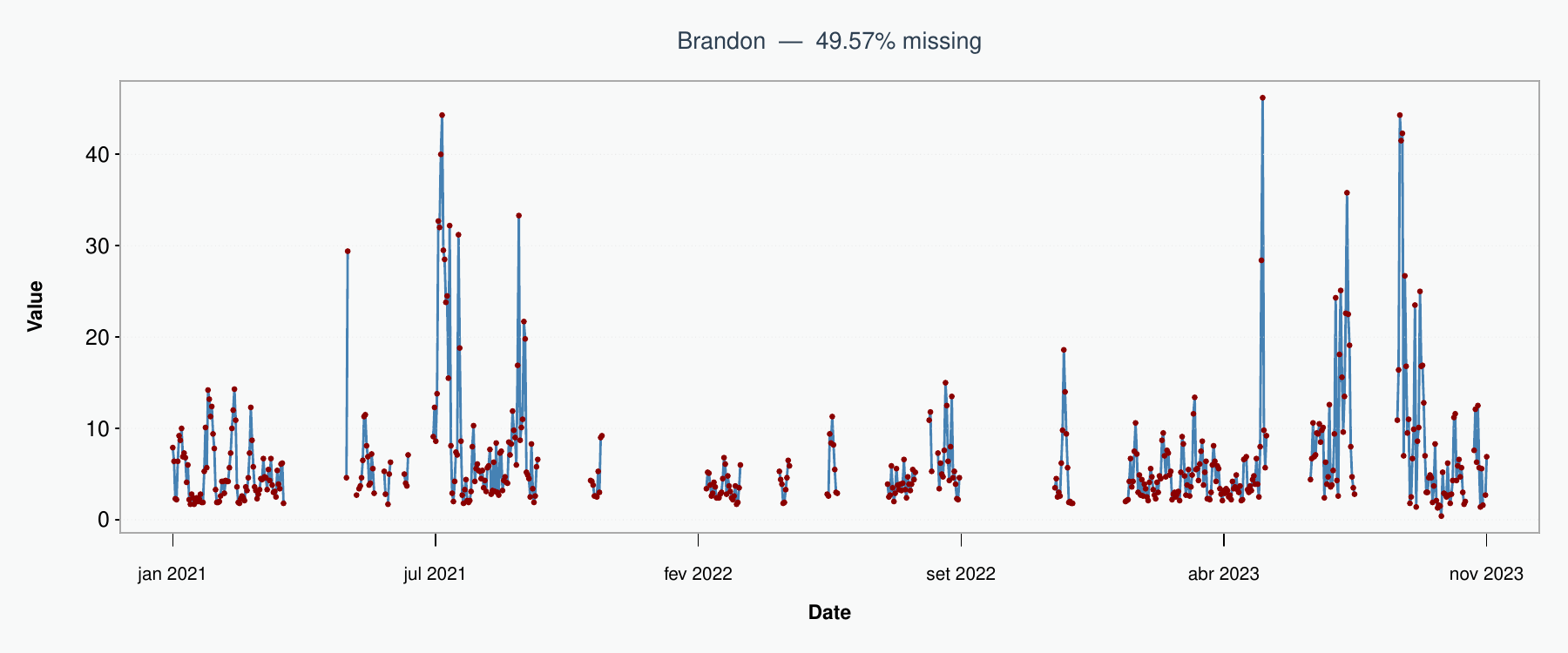}}
    \mbox{
    \includegraphics[width = 0.8\textwidth]{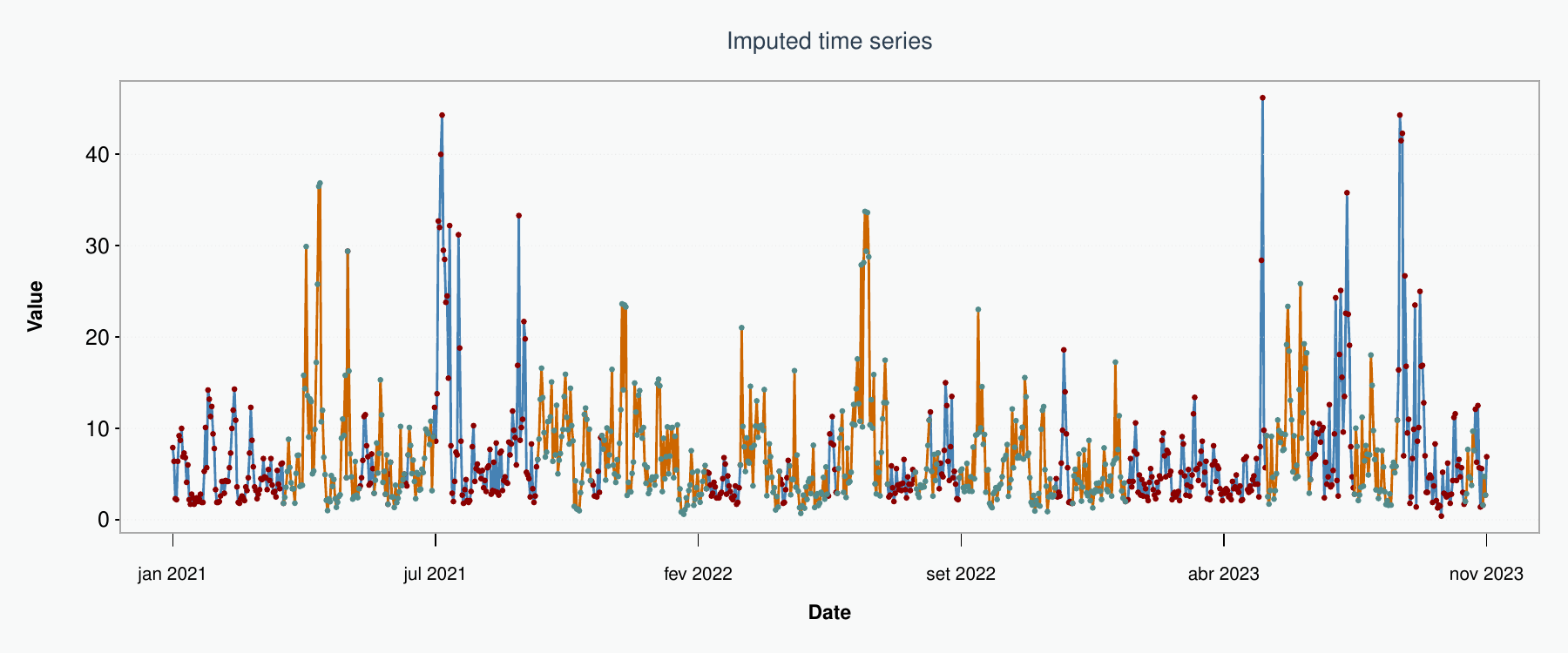}}
    \caption{Time-series plot - (top) daily average concentration of PM${}_{2.5}$ from the city of Brandon; (bottom) Imputed data in the last step of Algorithm 1.}
    \label{fattempts}
\end{figure}

Time series plots showing the missing and observed data are presented in the top panel of Figure~\ref{fattempts}. Since the data are strictly positive, Gaussian models are not adequate, and we instead consider a observation-driven model (ODM) based on Gamma-GARMA models. This model assumes that the random component follows $Y_t|\mathcal{F}_{t-1}\sim \mathrm{Gamma}(\mu_t,\nu)$, a Gamma distribution parametrized by its mean, while the systematic component follows \eqref{arma}. All analyses were performed in \texttt{R} using the \texttt{BTSR} package.

Starting values were selected according to Section~\ref{sv}, yielding a slice of  size 149. Following a Box--Jenkins identification procedure, our first attempt specified a Gamma-GARMA$(1,1)$ model with initial parameters $\hat{\gamma}_0$ (Table~\ref{attempts}), all statistically significant. Increasing the order $p$ or $q$ yielded models with non-significant coefficients.

Using this $\hat{\gamma}_0$, we applied Algorithm~1 with $K=30$, employing both stopping criteria from Section~\ref{sc} with threshold $\tau=0.01$. The results reveal an interesting pattern: while CVSC required 10 iterations to converge,  VRSC converged in only 2 steps, with similar results (Table~\ref{attempts}, 1st Attempt). The near-zero estimates for $\theta$ suggest that the MA component may be redundant. This outcome is not entirely surprising given that the  initial estimates $\hat{\gamma}_0$ were obtained from a relatively small  starting slice of only 149 observations. With limited data, the initial specification may have overparameterized the model, capturing noise rather than genuine dynamics. This point motivate the application of a reduced Gamma-GARMA$(1,0)$ specification.

The second attempt confirms this intuition. Starting from $\hbs{\gamma}_0 = (0.639, 0.689, 4.044)$, both stopping criteria now converge to remarkably similar estimates within a comparable number of iterations (4-5 steps). The autoregressive parameter $\hat{\phi} \approx 0.645$ indicates moderate  persistence in PM${}_{2.5}$ concentrations, with shocks decaying gradually rather than abruptly. The dispersion parameter $\hat{\nu} \approx 3.4$ suggests the data exhibit less variability than an exponential distribution (which would  correspond to $\nu = 1$), consistent with the bounded nature of the observed concentrations.

The convergence behavior across attempts is also instructive. In the first attempt, VRSC's rapid convergence (2 iterations) appears to have terminated prematurely. In contrast, the second attempt shows both criteria converging to nearly identical values, suggesting that the reduced model specification leads to more stable optimization. The imputed time series obtained from the last iteration of the second attempt (bottom panel of Figure~\ref{fattempts}) appears visually consistent with the observed data patterns.

\FloatBarrier

\begin{table}[!ht]
\caption{Results related to first and second attempts of analyzing the PM${}_{2.5}$ data.}\label{attempts}\vspace{.3cm}
\centering
\setlength{\tabcolsep}{3pt}
\renewcommand{\arraystretch}{1.2}
\small
\begin{tabular}{c||ccccc||cccc}
\hline
 &\multicolumn{5}{c||}{1st Attempt}&\multicolumn{4}{c}{2nd Attempt}\\
\hline
Method & $\alpha$ & $\phi$ & $\theta$ & $\nu$& iter & $\alpha$ & $\phi$ & $\nu$ & iter\\
\hline
 $\hbs\gamma_0$ & 1.069 & 0.449 & 0.410 & 4.231 & -- & 0.639 & 0.689 &  4.044 & --\\
 \hdashline
 CVSC & 0.729 & 0.645 & -0.005 & 3.389 & 10  & 0.731 & 0.645 & 3.443 & 5\\
VRSC & 0.775 & 0.620 &   0.055 & 3.433 &  2 &0.730 & 0.646 &  3.387 & 4\\
\hline
\end{tabular}
\end{table}

\subsection*{Acknowledgments}
G. Pumi and T.S. Prass gratefully acknowledge the financial support received by the Conselho Nacional de Desenvolvimento Cient\'ifico e Tecnol\'ogico – Brasil (CNPq) – Bolsa de Produtividade em Pesquisa - Proc. 303281/2025-1 (Pumi) and 305886/2025-8 (Prass).
\subsection*{Conflict of Interest Statement}
The author declares no conflicts of interest.
\subsection*{Data Availability Statement}
The data that support the findings of this study are openly available. The data considered in Section \ref{stored} are available from Operador Nacional do Sistema El\'etrico in Brazil, {\color{blue}\small\url{https://www.ons.org.br/Paginas/resultados-da-operacao/historico-da-operacao/energia_armazenada.aspx}}. The data considered in Section \ref{pol} are available from the  United States Environmental Protection Agency AirNow program's website {\color{blue}\small\url{https://docs.airnowapi.org/}}.
\bibliographystyle{apalike}
\bibliography{bib_proj}

\includepdf[pages=-]{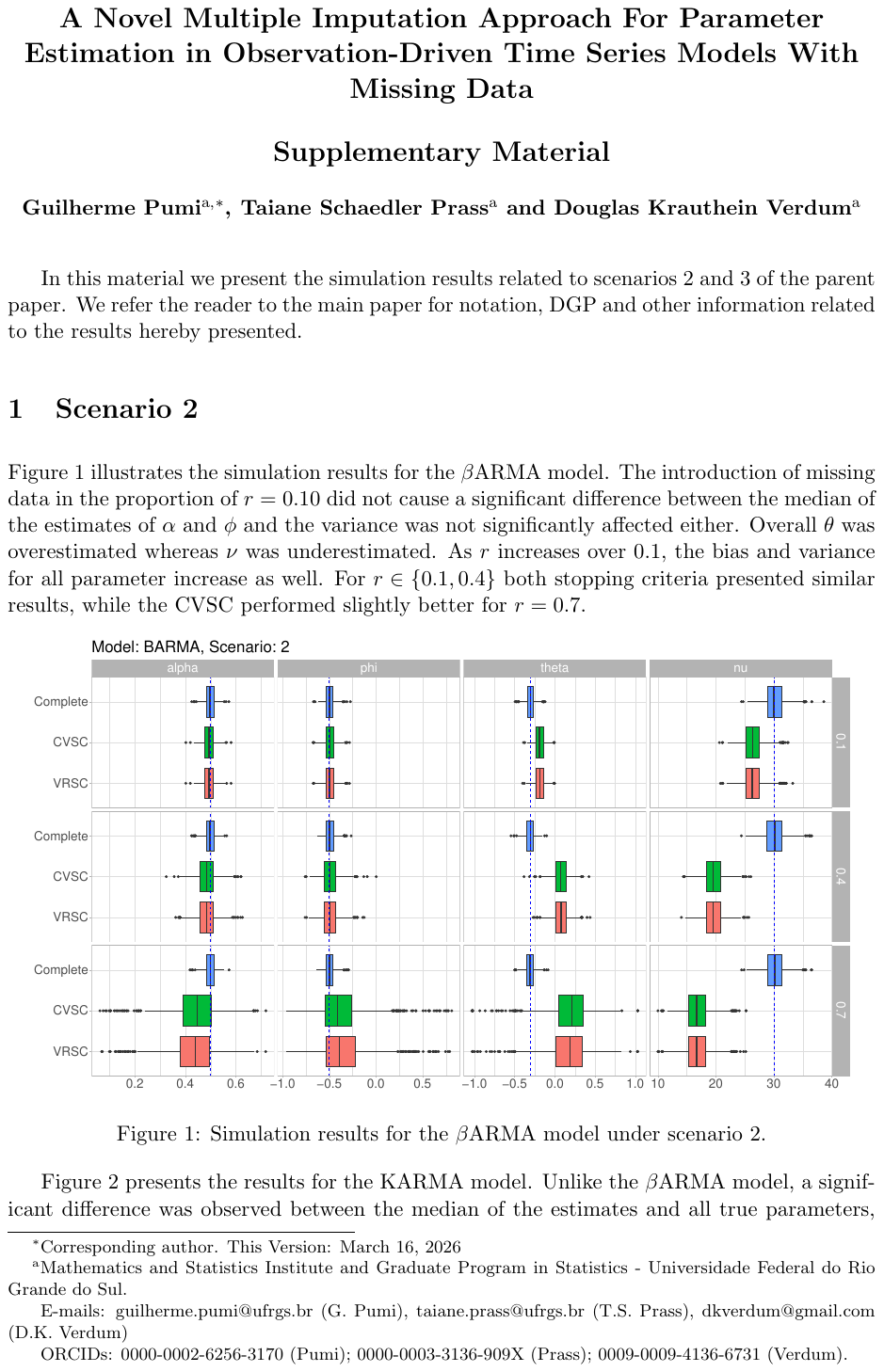}

\end{document}